\documentclass[sigconf,natbib=true]{acmart}
\renewcommand\footnotetextcopyrightpermission[1]{}
\usepackage{amsmath}
\usepackage{algorithm}
\usepackage{algorithmic}
\usepackage{multirow}
\usepackage{xspace}
\usepackage{subfigure}
\usepackage{enumitem}
\usepackage{makecell}

\newcommand{\VAE}{DQ-VAE\xspace}
\newcommand{\method}{DQRec\xspace}

\AtBeginDocument{%
  }

\setcopyright{acmlicensed}
\copyrightyear{2025}
\acmYear{2025}
\acmDOI{XXXXXXX.XXXXXXX}
\acmConference[Conference acronym 'XX]{Make sure to enter the correct
  conference title from your rights confirmation email}{June 03--05,
  2018}{Woodstock, NY}
\acmISBN{978-1-4503-XXXX-X/18/06}

\begin{document}

\title{Representation Quantization for Collaborative Filtering Augmentation
}


\author{Yunze Luo}
\affiliation{
    \institution{School of CS, Peking University}
    \city{Beijing}
    \country{China}}
\email{lyztangent@pku.edu.cn}

\author{Yinjie Jiang}
\affiliation{
    \institution{Kuaishou Technology}
    \city{Beijing}
    \country{China}}
\email{jiangyinjie@kuaishou.com}

\author{Gaode Chen}
\authornote{Corresponding author.}
\affiliation{
    \institution{Kuaishou Technology}
    \city{Beijing}
    \country{China}}
\email{chengaode19@gmail.com}

\author{Jingchi Wang}
\affiliation{
    \institution{School of CS, Peking University}
    \city{Beijing}
    \country{China}}
\email{jamesw12@stu.pku.edu.cn}

\author{Shicheng Wang}
\affiliation{
    \institution{Kuaishou Technology}
    \city{Beijing}
    \country{China}}
\email{wangshicheng@kuaishou.com}

\author{Ruina Sun}
\affiliation{
    \institution{Kuaishou Technology}
    \city{Beijing}
    \country{China}}
\email{srn672836152@163.com}

\author{Yuezihan Jiang}
\affiliation{
    \institution{Unaffiliated}
    \city{Beijing}
    \country{China}}
\email{yuezihan.jiang@gmail.com}

\author{Jun Zhang}
\affiliation{
    \institution{Kuaishou Technology}
    \city{Beijing}
    \country{China}}
\email{zhangjun08@kuaishou.com}

\author{Jian Liang}
\affiliation{
    \institution{Kuaishou Technology}
    \city{Beijing}
    \country{China}}
\email{liangjian03@kuaishou.com}

\author{Han Li}
\affiliation{
    \institution{Kuaishou Technology}
    \city{Beijing}
    \country{China}}
\email{lihan08@kuaishou.com}

\author{Kun Gai}
\affiliation{
    \institution{Kuaishou Technology}
    \city{Beijing}
    \country{China}}
\email{yuyue06@kuaishou.com}

\author{Kaigui Bian}
\authornotemark[1]
\affiliation{
    \institution{School of CS, Peking University}
    \city{Beijing}
    \country{China}}
\email{bkg@pku.edu.cn}

\thanks{Yunze Luo, Jingchi Wang, and Kaigui Bian are affiliated with School of CS, AI Innovation Center, National Engineering Laboratory for Big Data Analysis and Applications, State Key Laboratory of Multimedia Information Processing, Peking University.}
\renewcommand{\shortauthors}{Yunze Luo et al.}

\begin{abstract}
As the core algorithm in recommendation systems, collaborative filtering (CF) algorithms inevitably face the problem of data sparsity. Since CF captures similar users and items for recommendations, it is effective to augment the lacking user-user and item-item homogeneous linkages. However, existing methods are typically limited to connecting through overlapping interacted neighbors or similar attributes and contents. These approaches are constrained by coarse-grained, sparse attributes and fail to effectively extract behavioral characteristics jointly from interaction sequences and attributes. To address these challenges, we propose a novel two-stage collaborative recommendation algorithm, \underline{DQRec}: Decomposition-based Quantized Variational AutoEncoder (DQ-VAE) for Recommendation. DQRec augments features and homogeneous linkages by extracting the behavior characteristics jointly from interaction sequences and attributes, namely \textit{patterns}, such as user multi-aspect interests. Inspired by vector quantization (VQ) technology, we propose a new VQ algorithm, DQ-VAE, which decomposes the pre-trained representation embeddings into distinct dimensions, and quantizes them to generate semantic IDs. We utilize the generated semantic IDs as the extracted \textit{patterns} mentioned above. By integrating these patterns into the recommendation through feature and linkage augmentation, the system enriches both latent and explicit features, identifies pattern-similar neighbors, and thereby improves the efficiency of information diffusion. Experimental comparisons with baselines across multiple datasets demonstrate the superior performance of the proposed DQRec method.

\end{abstract}

\ccsdesc[500]{Information systems~Recommender systems}

\keywords{Recommender System, Collaborative Filtering, Vector Quantization, Semantic IDs}


\maketitle

\section{Introduction}
With the rise of various e-platforms, recommender systems are playing an increasingly pivotal role in people's lives by helping users explore items~\citep{MGL, PAM}.
Collaborative filtering (CF)~\cite{CF, NCF} is one of the most commonly used methods in recommender systems, which globally optimizes the network parameters and embeddings by identifying commonalities in the interactions between users and items. Recently, graph neural networks (GNNs)~\cite{NGCF, LightGCN} and sequential recommendation~\cite{GRU4Rec, SASRec} have demonstrated a powerful ability to extract collaborative information. Graph-based CF can learn meaningful representations by leveraging the interaction data between users and items, and sequential recommendation focuses on modeling users' interaction sequences, aiming to capture users' dynamic interests and predict the next interacting item.

Existing collaborative filtering methods have already shown excellent performance. However, we observe that those methods still face the following challenges. 

The goal of collaborative filtering methods is to capture the commonalities between similar users and items based on the user-item interaction information and to predict future user-item interactions as accurately as possible~\cite{NCF, CF}. Therefore, the efficiency of CF algorithms is heavily impacted by the density of user-item interaction data~\cite{Sparsity, A-LLM}.
However, the user-item interaction data is extremely sparse, which often leads to suboptimal efficiency in CF algorithms. Since CF captures similar users and items for recommendations, it is effective to \textbf{augment the lacking user-user and item-item homogeneous linkages}~\cite{SHaRe}. By augmenting these homogeneous linkages, information diffusion in similar users and items can be directly performed, alleviating the data sparsity issue in CF.

To address the augmentation of homogeneous linkages, a common approach is to utilize existing historical interaction data to establish linkages (\textit{e.g.}, two-hop neighbors)~\cite{HCCF, LightGCN, GIFT}. 
However, such methods face challenges as they fail to capture behavior patterns from user or item interaction sequences to measure the similarity between users or items and establish homogeneous links accordingly.
For example, two users might browse different sequences of items, but their interests could lie in the same category of content, thus exhibiting similar behavioral interests and deserving to be connected.
Another approach focuses on establishing linkages based on similar attributes~\cite{MGL, M3CSR}. These methods have two main challenges:
\textit{(i)} Lack of fine-grained attributes. Existing datasets often lack detailed user attributes~\cite{user-side-info}, typically only containing basic information such as gender and age, which makes it difficult to explore homogeneous linkages for users.
\textit{(ii)} Bias between attribute and behavior. 
Focusing solely on attributes ignores the information contained in the interaction sequences, preventing the joint measurement of the potential characteristics among attributes and behaviors.
Therefore, existing collaborative filtering enhancement approaches still face challenges.

Recognizing the aforementioned challenges, the question arises --- how can we effectively model behavioral characteristics, namely \textit{patterns}, by jointly leveraging both the attributes and behaviors of users and items and utilizing the \textit{patterns} to augment collaborative filtering for optimizing recommendation performance?
Since extracting common patterns by interactions and attributes can be seen as a form of aggregation, we employ vector quantization (VQ)~\cite{VQ-VAE} technique to generate semantic quantized encodings as \textit{pattern} for users and items, and explore the latent diffusion relationships between users and between items. In recent years, VQ technology has demonstrated its superiority across an increasing number of fields, including image generation~\cite{VQ-VAE2} and collaborative recommendation~\cite{VQ-Rec}. VQ can transform vectors that require quantization through an encoder and generate their semantic IDs. Existing VQ techniques for recommendation~\cite{TIGER} encode item embeddings, which are generated from side-information, into semantic IDs. These semantic IDs can be considered efficient features of item commonalities and can be utilized to discover potential diffusion signals through similar encoding IDs. However, such fixed encodings fail to reflect the pattern of the interaction sequence of items and cannot generate semantic IDs for users to capture their interests. Therefore, it is necessary to design unique systems that capture the pattern of both users and items.
For example, the generated pattern sequence of a user may capture multi-aspects of the user’s interests, such as a preference for yellow, expensive, and photography-related items.
By leveraging the generated patterns, we can enhance linkages to supplement sparse homogeneous linkages.

In this work, we propose a novel two-stage framework named \underline{\method}: \textbf{D}ecomposition-based \textbf{Q}uantized Variational AutoEncoder (DQ-VAE) for \textbf{Rec}ommendation, which effectively extracts patterns of users and items to enhance collaborative recommendation. First, a pre-trained model that encodes the features and interaction sequences of users and items into corresponding representation embeddings, and
the proposed \VAE technique is applied to quantize the representations embeddings of users and items, extracting multi-dimensional, disentangled semantic ID sequences as \textit{patterns}. Ultimately, by leveraging the generated semantic ID patterns, we enhance collaborative filtering from two perspectives: \textit{(i)} We directly utilize the semantic ID patterns as attributes to capture commonalities among users and items, thereby enriching their features. \textit{(ii)} We propose a unique approach for enhancing collaborative information by mining both explicit and latent patterns, thereby strengthening the homogeneous linkages and facilitating better information diffusion. Through the aforementioned approach, we enhance the efficiency of information diffusion in collaborative filtering, addressing the challenges outlined earlier.

Our main contributions can be summarized as follows:

\begin{itemize}[leftmargin=*]
\item We effectively tackle the challenges present in existing homogeneous linkage augmentation methods by jointly extracting \textit{patterns} from interactions and attributes, therefore alleviating the performance degradation caused by data sparsity in CF.
\item We propose a novel \VAE paradigm that efficiently quantizes the representations of users and items into semantic ID encodings, extracting multi-aspect patterns of behavior.
\item We conduct a comprehensive experiment on three public datasets, demonstrating performance improvement over various baselines, and highlighting the improvements brought by each component. 
\end{itemize}

\section{Related Works}
In this section, we discuss research relevant to this work, including general collaborative filtering and vector quantization.

\subsection{Collaborative Filtering}

Collaborative filtering (CF)~\cite{CF} algorithms represent a milestone in recommendation systems, focusing on using historical user-item interaction data to predict potential future interactions. The main idea of CF is to generate representations for users and items and use their similarity to make recommendations~\cite{NCF}. In recent years, GNN-based collaborative filtering has gained prominence as an evolving approach~\cite{NGCF, LightGCN, GMCF, StarGCN, HCCF, SimRec, LLMRec}. The fundamental idea is to treat user-item interactions as a bipartite graph, propagating information between users and items through the graph’s links and optimizing the performance of similar nodes based on their interaction histories. Sequential recommendation is another common collaborative filtering approach~\cite{Bert4Rec, SASRec, CASER, ICRec, DuoRec}, which models user representations based on their historical behavior sequences, with many methods integrating GNNs for enhanced modeling~\cite{SRGNN, MAERec}. However, while these algorithms effectively capture user and item representations, they are still constrained by the sparsity of homogeneous links between users and between items, while failing to extract the potential patterns of users and items.

\subsection{Vector Quantization}

Since the introduction of Vector Quantized Variational AutoEncoder (VQ-VAE)~\cite{VQ-VAE} for image encoding~\cite{VQ-VAE2}, this technique has found widespread application across various domains. Starting from PQ-VAE\cite{PQVAE}, VQ-VAE began to demonstrate its potential in recommender systems. VQ-Rec~\cite{VQ-Rec} employed VQ-VAE to generate quantized encodings for items, facilitating modality alignment. Another emerging VQ technique RQ-VAE~\cite{RQ-VAE} has gained attention due to its hierarchical structure, making it suitable for item quantized encoding~\cite{TIGER, SID2, LCREC}. The generated item semantic ID sequences can be directly applied to generative recommendation tasks~\cite{TIGER} or used as inputs to large language models to perform recommendation directly~\cite{LCREC}.
However, the aforementioned methods fail to effectively generate quantized encodings for users and do not incorporate interaction sequences to capture patterns.


\section{Preliminary}
\subsection{Notations}
In this work, we consider the general CF scenario. Let $\mathcal{U}=\{u_1, \cdots,$ $ u_U\}$ and $\mathcal{I}=\{i_1, \cdots, i_I\}$ denote the universal set of users and items, and $U$ and $I$ represent the number of users and items. $\mathcal{S}_u=\{i_1^u, \cdots, i_{|\mathcal{S}_u|}^u\}$ denotes the recent interaction sequence of user $u$, and $\mathcal{S}_i$ is similarly the sequence of item $i$. Compared to the interaction matrix $\boldsymbol{R}\in\{0,1\}^{U\times I}$, the recent interaction sequence provides a more effective representation of the characteristics of users and items. $(\mathcal{T}_u,\mathcal{T}_i,y_{ui})=((u, \mathcal{S}_u), (i,\mathcal{S}_i), y_{ui})\in\mathcal{N}$ construct a interaction sample in dataset $\mathcal{N}$, where $y_{ui}$ indicates the ground truth of interaction. Matrices and vectors are denoted by bold symbols.

\subsection{Problem Formulation}
The objective of CF recommendation systems is to predict interaction labels as accurately as possible based on the existing information about users and items. For a given sample $(\mathcal{T}_u, \mathcal{T}_i, y_{ui})\in \mathcal{N}_\mathrm{train}$, the system takes the interaction sequence, and attribute information of user and item as inputs, and outputs the predicted results for their interaction. The system is optimized as follows:
\begin{equation}
    \arg\min_{\Theta} \sum_{(\mathcal{T}_u, \mathcal{T}_i, y_{ui})\in \mathcal{N}_\mathrm{train}}\mathcal{L}\left(\mathcal{W}^{\Theta}(\mathcal{T}_u, \mathcal{T}_i), y_{ui}\right)
\end{equation}
where $\mathcal{W}$ represents the prediction recommender regarding to learnable parameters $\Theta$, and $\mathcal{L}$ denotes the loss function.




\section{Methods}

\subsection{Overview}
In this section, we will provide a detailed description of our proposed approach \method.

First, the features of users and items, along with the features of neighbors in the historical interaction sequences, are encoded into representation embeddings by a pre-trained model.
The proposed \VAE is then trained to quantize the representation embedding and generate semantic IDs as patterns. Users and items each have their independent autoencoders. Taking users as an example, the system generates the encoder for each layer of the VAE based on the stored representation embeddings. Subsequently, the system optimizes the embeddings of each layer’s codebook through training. Once the training is complete, the parameters of the VAE remain frozen during the training of the recommendation model.

After the training of \VAE, the system trains the recommendation model on the training set. For each input user-item pair, the semantic ID pattern is calculated using the embedding encoding model and \VAE, and the semantic ID pattern is then utilized in training the recommendation system, including feature augmentation and linkage augmentation.

\subsection{Embedding Generation and \VAE}
\begin{figure*}[t]
    \centering
    \includegraphics[width=\linewidth]{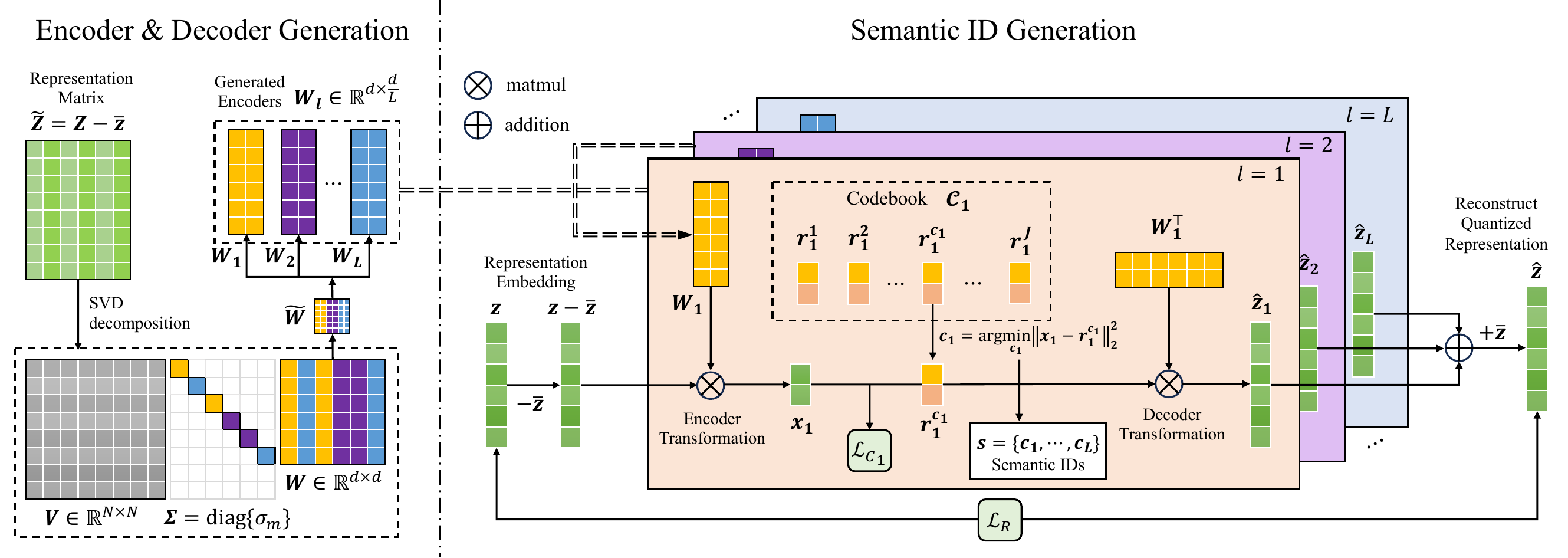}
    \caption{The structure of the proposed \VAE framework. (1) The encoders and decoders of DQ-VAE are generated by SVD to equally divide the information in the representations into dimensions. (2) Semantic IDs and the quantized representations are calculated through DQ-VAE, by calculating nearest embeddings in different layers of codebooks.}
    \label{fig:VAE}
\end{figure*}

In this section, we will describe how the designed \VAE generates semantic ID patterns. Fig.~\ref{fig:VAE} illustrates the structure of \VAE in detail. We define the semantic IDs as a sequence of $L$ codewords. 
Ideally, taking users as an example, more similar semantic IDs indicate more similar patterns.
Ideally, each position in the ID sequence corresponds to a distinct feature dimension, and identical IDs at a particular position indicate similar features along that dimension.

Since the semantic ID is a sequence, each ID in the sequence represents a different dimension of the pattern, such as a user’s preferences for different attributes of an item. In our scenario, we aim for the generated IDs to exhibit decoupling across dimensions, ensuring that different interest dimensions remain independent of one another. This allows us to model patterns effectively, enabling the exploration of one pattern dimension without affecting others. 
Currently, many vector quantization-based approaches~\cite{TIGER, LC-REC} have been applied to recommendation systems, where semantic IDs are generated for items based on their attributes to facilitate recommendations. However, the RQ-VAE~\cite{RQ-VAE} technique commonly used in these methods does not meet the requirements of our scenario. In the hierarchical generation process of RQ-VAE, the ID calculation of each layer depends on the IDs from the previous layer. This introduces correlations between IDs at different layers, where exploration at one layer can influence the generation of IDs at other layers. Moreover, as a hierarchical structure, it lacks sufficient interpretability when IDs at higher layers differ while sharing identical IDs at lower layers. 

In our \VAE's design, we leverage singular value decomposition~\cite{SVD} (SVD) of matrices to identify a linear transformation that converts the representation embedding into a form where dimensions are mutually independent. Furthermore, this transformation is decomposed into multiple components to represent different dimensions of the representation. In this manner, each part of the linear transformation can be viewed as an encoder, and the transformed vector is subsequently quantized to produce the semantic ID for the corresponding dimension.

\subsubsection{Embedding Generation}
First, a pre-trained model is applied to generate the embeddings for users and items. These embeddings will later be fed into \VAE for training. This pre-trained model can either be a language model (\textit{e.g.} BERT~\cite{BERT}) that contains world knowledge or a standard recommendation model. The key idea behind generating the embeddings is that we input only the features of users and items, along with the features of their neighbors in the interaction sequences, without using their IDs. The purpose of this approach is to maintain the generalization performance of the model and better optimize the representations of different features, rather than focusing on a specific user or item.

Since the pre-trained model does not impose any specific requirements on the types of input features, any form of feature can be encoded and subsequently trained, thus avoiding the issue of dataset heterogeneity. In our design, we simply adopted a dual-tower~\cite{dual-tower} recommendation model and used the top embeddings as the generated representation embeddings.

\subsubsection{Encoder \& Decoder Calculation via SVD}
To extract the information embedded in different dimensions of representation embeddings, inspired by the matrix dimension reduction method PCA~\cite{PCA}, we adopt the SVD approach. Since the embeddings of users and items encapsulate distinct types of information, we process users and items separately and design dedicated autoencoders for each. Taking users as an example. Consider all users in the training set, whose representation embeddings $\boldsymbol{z}\in\mathbb{R}^d$ can form a user representation matrix $\boldsymbol{Z}\in\mathbb{R}^{N\times d}$, where $d$ denotes the embedding dimension of representation, $N$ is the number of users in training set, and assume that $N>d$.

In the trained representation embeddings, there may be correlations between dimensions, making it challenging to encode semantic ID patterns effectively. To address this, we aim to identify a linear transformation that can decorrelate the dimensions. Based on this transformation, we design the autoencoder to achieve the desired encoding. First, we need to centralize the matrix to ensure that the mean of the data in each dimension is zero, facilitating the subsequent variance computation:
\begin{equation}
    \tilde{\boldsymbol{Z}} = \boldsymbol{Z} - \bar{\boldsymbol{z}} = \boldsymbol{Z} - \frac{1}{N}\sum_{u=1}^{N}\boldsymbol{z}
    \label{equ:mean}
\end{equation}
and the mean of the training set’s representation embeddings $\bar{\boldsymbol{z}}$ is stored and used for subsequent training of the autoencoder.

To find such a linear transformation, we employ the SVD method for the centralized data matrix $\tilde{\boldsymbol{Z}}$:
\begin{equation}
    \tilde{\boldsymbol{Z}} = \boldsymbol{V}\boldsymbol{\Sigma}\boldsymbol{W}^\top
\end{equation}
where $\boldsymbol{V}\in\mathbb{R}^{N\times N}$, $\boldsymbol{W}\in\mathbb{R}^{d\times d}$ are matrices with orthogonal unit vectors as columns, and $\boldsymbol{\Sigma}\in\mathbb{R}^{N\times d}$ is a rectangular diagonal matrix containing the singular values of $\boldsymbol{Z}$. Here, we use $\boldsymbol{W}$ as the target linear transformation:
\begin{equation}
    \boldsymbol{Y} = \tilde{\boldsymbol{Z}}\boldsymbol{W}=\boldsymbol{V}\boldsymbol{\Sigma}
\end{equation}
where $\boldsymbol{Y}\in\mathbb{R}^{N\times d}$ is the transformed data.

Consider the covariance matrix of the transformed data:
\begin{equation}
    \mathrm{Cov}(\boldsymbol{Y}) = \frac{1}{N}\boldsymbol{Y}^\top\boldsymbol{Y}=\frac{1}{N}\boldsymbol{\Sigma}^\top\boldsymbol{V}^\top\boldsymbol{V}\boldsymbol{\Sigma}=\frac{1}{N}\hat{\boldsymbol{\Sigma}}^2
\end{equation}
where $\hat{\boldsymbol{\Sigma}}=\mathrm{diag}(\sigma_1, \cdots, \sigma_d)\in\mathbb{R}^{d\times d}$ represent the chopped square diagonal matrix with singular values $\sigma_n$ as diagonals. It can be observed that in the transformed data, the covariance between different dimensions is zero, satisfying the requirement for decorrelation across dimensions.

We use the variance of each dimension to represent the amount of information contained in that dimension. Since the diagonal elements of the covariance matrix represent the variance within each dimension, we establish a relationship between the variance of a dimension and its singular value: 
\begin{equation}
    \mathrm{D}(\boldsymbol{y}_n)=\frac{1}{N}\sigma_n^2\propto\sigma_n^2
\end{equation}
where $\boldsymbol{y}_n$ is the $n$-th column of $\boldsymbol{Y}$ and $\mathrm{D}$ represents the variance. 

Our designed \VAE consists of multiple parallel layers $L$, each comprising an encoder $\mathcal{E}_l$ and a codebook $\mathcal{C}_l$ including multiple embeddings. For an input representation embedding, each layer of the autoencoder encodes it through the encoder and retrieves the semantic ID of that layer $c_l$ via the codebook. Finally, the semantic IDs generated from all layers are concatenated to produce the quantized results $s$, and the quantized representation can be reconstructed through a decoder $\mathcal{D}$. In our design, both the encoder and decoder perform linear transformations, and we utilize the linear transformation $\boldsymbol{W}$ computed as described above to generate the encoder for each layer and the decoder.

With the linear transformation $\boldsymbol{W}$ computed as described above, we can map the original data into decorrelated data across different dimensions. 
Our goal is to divide $\boldsymbol{W}$ into $L$ submatrices along the columns so that the representation embeddings can be transformed by these submatrices into mutually independent components. When assigning columns to form submatrices, we aim to ensure that the amount of information contained in each submatrix is as similar as possible. Specifically, the sum of the differences in the squared singular values of each submatrix is minimized:
\begin{equation}
    \tilde{\boldsymbol{W}} = [\boldsymbol{W}_1, \cdots, \boldsymbol{W}_L]=\arg\min_{\tilde{\boldsymbol{W}}}\sum_{l_1\neq l_2}\left(\sum_{\boldsymbol{w}_m\in \boldsymbol{W}_{l_1}}\sigma_m^2-\sum_{\boldsymbol{w}_m\in \boldsymbol{W}_{l_2}}\sigma_m^2\right)
\end{equation}
where $\tilde{\boldsymbol{W}}$ is the column permutation of $\boldsymbol{W}$, and $\boldsymbol{W}_l\in\mathbb{R}^{d\times \frac{d}{L}}$ denotes the $l$-th encoder transformation that containing $\frac{d}{L}$ columns. 

Specifically, since the square of each singular value $\sigma_n^2$ corresponds to the amount of information contained in its associated dimension $\boldsymbol{y}_n$, we group the columns of $\boldsymbol{W}_l$ such that each group maintains a roughly equal amount of information (\textit{i.e.}, the sum of the squared singular values in groups is approximately balanced). This ensures that each layer of the semantic IDs contains a similar amount of information.

\subsubsection{Variantional Autoencoder}
Next, we detail the quantization process of the autoencoder. For the input representation embedding $\boldsymbol{z}$, the encoder at each layer operates on the centralized input to obtain the latent representation for that layer. For layer $l$:

\begin{equation}
    \boldsymbol{x}_l = \mathcal{E}_l(\boldsymbol{z}) = (\boldsymbol{z} - \bar{\boldsymbol{z}})\boldsymbol{W}_l
\end{equation}
where $\boldsymbol{x}_l\in\mathbb{R}^{\frac{d}{L}}$, $\boldsymbol{W}_l$ denotes the latent representation and encoder transformation of layer $l$, and $\bar{\boldsymbol{z}}$ is the mean stored in Eq.~(\ref{equ:mean}).

For each layer $l$, its codebook consists of several embeddings, denoted as $\mathcal{C}_l = \{\boldsymbol{r}_l^j\in\mathbb{R}^{\frac{d}{L}}\}_{j=1}^J$ , where $J$ represents the size of the codebook. The quantization process for the latent representation is achieved by finding the most similar embedding:

\begin{equation}
    c_l = \arg\min_{c_l} \|\boldsymbol{x}_l - \boldsymbol{r}_l^{c_l}\|^2
    \label{equ:codebook}
\end{equation}
where $c_l$ represent the $l$-th codeword in semantic IDs. By computing the latent representation and finding the most similar codebook embedding at each layer, we can obtain the quantized semantic ID sequence of the input $\boldsymbol{z}$: $s=\{c_1, \cdots, c_L\}$.

\begin{figure*}[t]
    \centering
    \includegraphics[width=\linewidth]{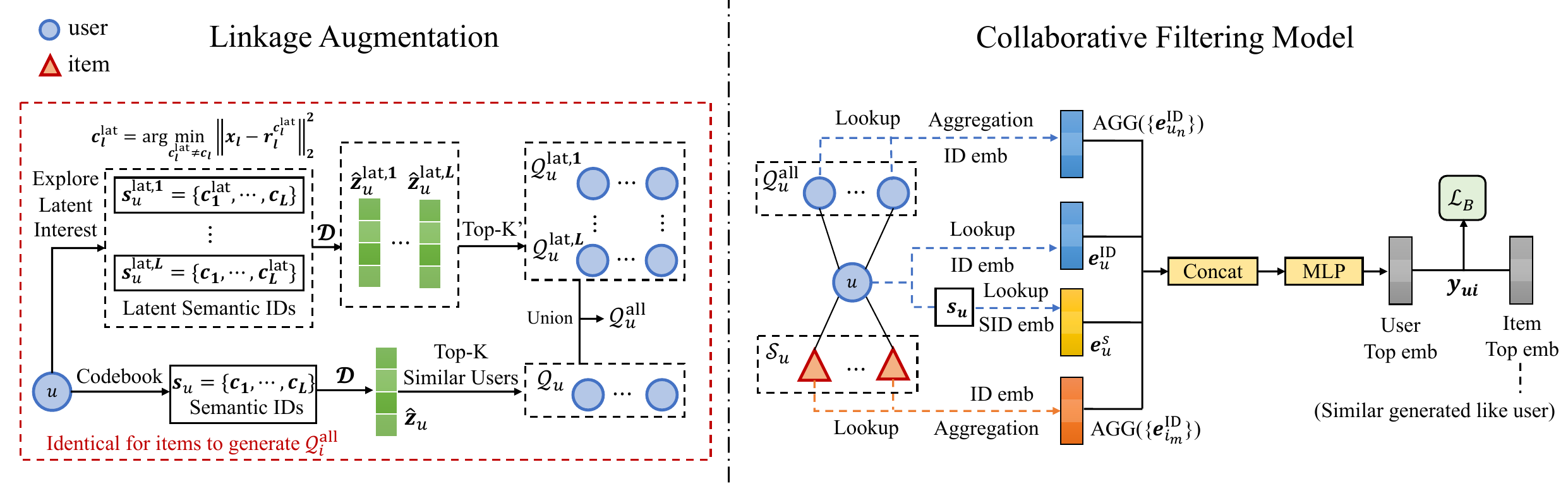}
    \caption{The detailed process of linkage augmentation and the collaborative filtering model. (1) Linkages are augmented through the semantic ID patterns. (2) Feature augmentation is performed by integrating patterns as features into recommendations.}
    \label{fig:CF}
\end{figure*}

After obtaining the semantic IDs, we can reconstruct the quantized representation embedding $\hat{\boldsymbol{z}}$. Since in the encoder, each column of $\boldsymbol{W}_l$ is an orthogonal unit vector, hence its transpose $\boldsymbol{W}_l^\top$ can be applied as the decoder. The reconstruction process is as follows:
\begin{equation}
    \hat{\boldsymbol{z}} = \mathcal{D}(s) = \sum_{l=1}^L \boldsymbol{r}_l^{c_l}\boldsymbol{W}_l^\top + \bar{\boldsymbol{z}}
\end{equation}

To train the \VAE, we design two loss functions following the process of VQ-VAE. First, the reconstructed representation should be as similar as possible to the input representation. This is achieved by minimizing the reconstruction loss:
\begin{equation}
    \mathcal{L}_R = \|\hat{\boldsymbol{z}}-\boldsymbol{z}\|_2^2
\end{equation}

In addition, the system aims to optimize the embeddings in the codebook to make them more evenly distributed across the feature space of each layer, improving the quantization of the representation embeddings of users or items. To achieve this, the system minimizes the distance between the latent representation and the closest quantized embedding, ensuring that the quantization process is effective. This loss can be expressed as:

\begin{equation}
    \mathcal{L}_C = \sum_{l=1}^L\|\boldsymbol{x}_l - \boldsymbol{r}_l^{c_l}\|_2^2
\end{equation}

Since the encoder of \VAE is fixed, there is no need for the stop gradient~\cite{VQ-VAE} operation similar to VQ-VAE. The overall loss function is therefore composed of two parts:

\begin{equation}
    \mathcal{L}_T = \mathcal{L}_R + \beta\mathcal{L}_C
\end{equation}
where $\beta$ is a hyperparameter controlling the weights of losses.



\subsection{Collaborative Filtering Augmentation}
In this section, we will describe how to use the extracted semantic ID patterns to enhance the feature, as well as to augment potential homogeneous linkages. The detailed process is shown in Fig.~\ref{fig:CF}.

\subsubsection{Feature augmentation}
After obtaining the semantic ID patterns for users and items, we can fully leverage the extracted representation information. The semantic IDs can serve as augmented features for both users and items, reflecting the patterns of users and items across multiple dimensions:

\begin{equation}
    \boldsymbol{e}_u^s = \left[\boldsymbol{F}_{U_1}^s({c_1}_u), \cdots, \boldsymbol{F}_{U_L}^s({c_L}_u)\right]
\end{equation}
where $\boldsymbol{F}_{U_l}^s$ represents the $l$-th semantic feature embedding matrix and $({c_l}_u)$ stands for taking the corresponding feature vector of the user from the matrix. Items have similar embedding $\boldsymbol{e}_i^s$ as users.

Through this approach, we obtain the pattern embeddings for both users and items.
The generated patterns offer two key advantages, dynamic attribute augmentation and the ability to capture commonalities.
\textit{(i)} Feature Augmentation. As mentioned above, user attributes are often sparse~\cite{LLMRec, user-side-info} and remain static within the system.
Pattern extraction addresses these issues by incorporating interest-shifting behaviors into the user’s representation, improving the ability to model the evolving nature of user preferences while augmenting high-quality features, and alleviating sparsity of attributes.
\textit{(ii)} Capturing Commonalities. The semantic ID patterns can also capture the commonalities between users and between items. If different users share the same semantic ID in a particular dimension, their pattern embeddings will share a portion of the same embedding, thereby enabling collaboration effects. 
By leveraging semantic IDs, the system can better portray group behavior and latent patterns of similarity, allowing for more effective information diffusion between users and between items.

\subsubsection{Linkage augmentation}
The generated patterns can also be applied to augment the potential homogeneous linkages between users and between items, 
thereby enhancing collaborative information.
In real-world recommendation systems, the input collaborative signals only include user-item interaction data, forming a bipartite graph, while lacking user-user and item-item linkages. Furthermore, since collaborative filtering aims to capture the similarity between nodes of the same type, the absence of these linkages significantly hampers the effectiveness of CF. 

In our design, the quantized representations $\hat{\boldsymbol{z}}$ are utilized to measure the similarities between users and between items. Take users as an example. During the training of the \VAE, we store the quantized representations of all users, denoted as $\hat{\mathcal{Z}} = \{\hat{\boldsymbol{z}}_1, \cdots, \hat{\boldsymbol{z}}_N\}$. Since users may appear multiple times in the training set, we store the quantized representation corresponding to the user’s most recent interaction in the timeline. Based on this list, for each sample in the training set $\mathcal{T}_u$, we can compute its semantic ID sequence $s_u$ and the corresponding quantized representation $\hat{\boldsymbol{z}}_u$. The representation is then compared with the stored representations of all users to calculate similarities, and the top-$K$ most similar users to the quantized representation are identified:

\begin{equation}
     \mathcal{Q}_u = \{u_1, \cdots, u_K\} = \mathrm{top}K\left(\{-\|\hat{\boldsymbol{z}}_u-\hat{\boldsymbol{z}}_n\|_2^2\}_{n=1}^N\right)
     \label{equ:topk}
\end{equation}
where $\mathrm{top}K$ representing selecting top $K$ users that minimize the distance. In this way, we use the quantized representations to find the $K$ users whose patterns are most similar to the current user’s. These users can then be used to enhance collaborative information and strengthen the representation of user $u$.

To further optimize collaborative filtering, we also design a unique pattern exploration mechanism. As mentioned earlier, a single pattern semantic ID in one dimension is often insufficient to effectively capture the potential pattern. To address this, by leveraging the quantifiable features in the embeddings of the VAE’s codebook, we can explore the potential of multiple patterns in a particular dimension. Similar to Eq.~(\ref{equ:codebook}), we can obtain the latent pattern of user $u$ in a particular dimension $l$ as follows:

\begin{equation}
    c_l^{\mathrm{lat}} = \arg\min_{c_l^{\mathrm{lat}}\ne c_l} \|\boldsymbol{x}_l - \boldsymbol{r}_l^{c_l^{\mathrm{lat}}}\|^2
\end{equation}
where $c_l^{\mathrm{lat}}$ represents the latent pattern of user. By replacing the explicit pattern ID $c_l$ with the latent pattern ID $c_l^{\mathrm{lat}}$, we can obtain the user’s latent pattern semantic ID:
\begin{equation}
    s^{\mathrm{lat}, l}=\{c_1,\cdots,c_l^{\mathrm{lat}, l},\cdots,c_L\}
\end{equation}
and quantized representation $\hat{\boldsymbol{z}}^{\mathrm{lat}, l}=\mathcal{D}(s^{\mathrm{lat}})$ in dimension $l$.

Following Eq.~(\ref{equ:topk}), we can similarly identify the top $K'$ most similar users based on the user’s quantized latent representation:
\begin{equation}
    \mathcal{Q}_u^{\mathrm{lat}, l} =  \{u_1, \cdots, u_{K'}\}
\end{equation}

By taking the union of the similar users based on the user’s latent patterns $\mathcal{Q}_u^{\mathrm{lat}, l}$ across all dimensions and those based on explicit patterns $\mathcal{Q}_u$, we can obtain the final set of similar users:

\begin{equation}
    \mathcal{Q}_u^{\mathrm{all}} = \mathcal{Q}_u\cup\left(\bigcup_{l=1}^L \mathcal{Q}_u^{\mathrm{lat}, l}\right)
\end{equation}

For items, we can similarly identify the similar neighboring items of item $i$ $\mathcal{Q}_i^{\mathrm{all}}$. Through the above process, we obtain similar neighbors of users and items via their quantized representations.

\subsubsection{Collaborative Filtering}
After obtaining the information that enhances collaborative filtering from both aspects, we can design the collaborative filtering recommendation system accordingly. In the specific design, we adopt a dual-tower~\cite{dual-tower} architecture.

The input embedding for users consists of several components:

\begin{equation}
    \boldsymbol{e}_u = \left[e_u^{\mathrm{ID}}, e_u^s, \mathrm{AGG}\left(\{\boldsymbol{e}_{i_m}^\mathrm{ID}\}_{i_m\in\mathcal{S}_u}\right), \mathrm{AGG}\left(\{\boldsymbol{e}_{u_n}^\mathrm{ID}\}_{u_n\in\mathcal{Q}_u^\mathrm{all}}\right)\right]
    \label{equ:input}
\end{equation}
where $\boldsymbol{e}_u$ is the input embedding, and $e_u^{\mathrm{ID}}$ represent the ID embedding of user $u$. $\mathrm{AGG}$ represents an aggregation function~\cite{LightGCN, BGNN, attention}.
In our work, we utilize the average pooling aggregator. Items have a comparable input embedding $\boldsymbol{e}_i$.


After the input embeddings are generated, they are fed into MLP and generate the top embeddings of users and items.
The embedding will be mapped by $L$ fully-connected network layers $\boldsymbol{z}=f_{H}\circ\cdots\circ f_{1}(\tilde{\boldsymbol{e}}^a)$, where the $h$-th layer can be represented as:
\begin{equation}
    f_{h}(\boldsymbol{e})=\phi(\boldsymbol{A}_h\boldsymbol{e}+\boldsymbol{b}_h)
\end{equation}
where $\boldsymbol{A}_h, \boldsymbol{b}_h$ are the weight matrix and bias vector of the $h$-th hidden layer, and $\phi$ represents a GELU~\cite{ReLU, GELU} activation function.

Finally,  the predicted scores for the corresponding user-item pairs are calculated based on the top embeddings of the user and item. The prediction of interaction $\hat{y_{ui}}$ is calculated as: 
\begin{equation}
     \hat{y}_{ui} = \sigma(\boldsymbol{z}_u \cdot \boldsymbol{z}_i)
\end{equation}
where $\sigma$ represents the sigmoid function for normalization.

We adopt the Bayesian Personalized Ranking~\cite{BPR}  loss function:

\begin{equation}
    \mathcal{L}_B = -\ln\sigma(\hat{y}_{ui}-\hat{y}_{ui^-})
\end{equation}
where $i^-$ is a negative sample that $(u, i^-)$ is unobserved. 


\section{Complexity Analysis}
Because of the additional introduction of the VAE model, \method will consume more resources than the base dual-tower model.

\textbf{Space Complexity.} The learned parameters in \method are from the user and item embeddings, parameters in \VAE codebooks and in dual-tower recommender model. The total number of parameters is $O\left((|\mathcal{U}|+|\mathcal{I}|)d+Ld+d^2\right)$, where $L$ is the number of codebook layers, with only the additional overhead from \VAE.

\textbf{Time Complexity.} Since the model is trained in two stages and the \VAE remains frozen after pretraining, the time and resource overhead during the training phase is acceptable. For online inference, the time complexity of semantic ID sequence generation and the dual-tower model are $O(Ld + d^2)$ and $O(|\mathcal{S}|d + Kd + d^2)$, respectively, given that \method uses average pooling to aggregate neighbor embeddings. The total complexity is therefore $O(Ld + |\mathcal{S}|d + Kd + d^2)$, where $K$ is the number of similar neighbors and $|\mathcal{S}|$ denotes the length of the
interaction sequences. If attention mechanisms were used for aggregation, as in traditional sequential recommendation methods, the complexity would become $O(Ld + |\mathcal{S}|^2d + Kd + |\mathcal{S}|d^2)$, which remains acceptable within an online inference framework.

\begin{table*}[t]
\centering
\caption{The reported top-K evaluation metrics. Bold represents the optimal and underlined represents the suboptimal results.}
\label{tab:perf}
\resizebox{\textwidth}{!}{%
{\small
\begin{tabular}{c|c|ccccccccccccc|c}
\toprule
Dataset & Metric & NCF & NGCF  &Bert4Rec & LGCN & HCCF  & DuoRec& CL4SRec&ICLRec & SimRec  &MAERec  & LLMRec  &VQRec& \method & Impr. \% \\ \midrule
\multirow{6}{*}{\textbf{MovieLens}} 
& Recall@5 &0.1285& 0.2018 &0.4652& 0.2342& 0.2495 & \underline{0.5277}& 0.5243&0.4518& 0.2460 &0.4919 &0.2569&0.4520&\textbf{0.6204}& +17.57\%\\
& Recall@10 &0.2186& 0.3092 &0.6060& 0.3500& 0.3718 & 0.6745& 0.6690&0.6180& 0.3667 &0.6492 &0.3773&\underline{0.6843}&\textbf{0.8333}& +21.77\%\\
& Recall@20 &0.3567& 0.4607 &0.7319& 0.5072& 0.5299 & 0.7944& 0.7930&0.7682& 0.5212 &0.7796 &0.5398&\underline{0.8770}&\textbf{0.9514}& +8.48\%\\
\cmidrule{2-16}&NDCG@5 &0.0807& 0.1379 &0.3365& 0.1595& 0.1706 & \underline{0.3837}& 0.3798&0.3135& 0.1681 &0.3444 &0.1776&0.3696&\textbf{0.4051}& +5.57\%\\
&NDCG@10 &0.1095& 0.1724 &0.3821& 0.1968& 0.2100 & 0.4314& 0.4267&0.3673& 0.2068 &0.3953 &0.2155&\underline{0.4414}&\textbf{0.4744}& +9.97\%\\
&NDCG@20 &0.1442& 0.2105 &0.4140& 0.2363& 0.2498 & 0.4619& 0.4582&0.4054& 0.2458 &0.4284 &0.2633&\underline{0.4772}&\textbf{0.5048}& +5.78\%\\
\midrule
\multirow{6}{*}{\textbf{Books}}& Recall@5 &0.2674& 0.3309 &0.3750& 0.3810& 0.3977 & 0.4026& \underline{0.4098}&0.2992& 0.4007 &0.3895 &0.3991 &0.3603&\textbf{0.4261}& +3.98\%\\
& Recall@10 &0.4226& 0.5117 &0.5316& 0.5578& 0.5576 & 0.5628& 0.5624&0.4584& \underline{0.5640} &0.5423 &0.5568 &0.5300&\textbf{0.5954}& +5.57\%\\
& Recall@20 &0.5981& 0.6919 &0.6860& 0.7094& 0.6978 & \underline{0.7151}& 0.7083&0.6280& 0.7140 &0.6904 &0.7135 &0.7006&\textbf{0.7494}& +4.80\%\\
\cmidrule{2-16}&NDCG@5 &0.1568& 0.1947 &0.2307& 0.2278& 0.2410 & \underline{0.2487}& 0.2457&0.1775& 0.2489 &0.2423 &0.2406 &0.2146&\textbf{0.2608}& +4.87\%\\
&NDCG@10 &0.2071& 0.2529 &0.2814& 0.2850& 0.2933 & \underline{0.3007}& 0.2955&0.2286& 0.2993 &0.2919 &0.2925 &0.2698&\textbf{0.3159}& +5.05\%\\
&NDCG@20 &0.2523& 0.3001 &0.3220& 0.3278& 0.3303 & \underline{0.3413}& 0.3345&0.2729& 0.3342 &0.3312 &0.3347 &0.3144&\textbf{0.3569}& +4.57\%\\
\midrule
\multirow{6}{*}{\textbf{Netflix}}
& Recall@5 &0.1507& 0.2405 &0.2997& 0.2638& 0.2435 & \underline{0.3378}& 0.3369&0.3065& 0.2863 &0.3231 &0.3079&0.3123&\textbf{0.4190}& +24.03\%\\
& Recall@10 &0.2507& 0.3516 &0.4175& 0.3821& 0.3586 & 0.4592& 0.4600&0.4298& 0.4114 &0.4472 &0.4285&\underline{0.5102}&\textbf{0.6505}& +27.50\%\\
& Recall@20 &0.3986& 0.4989 &0.5605& 0.5356& 0.5097 & 0.5962& 0.6013&0.5759& 0.5665 &0.5945 &0.5884&\underline{0.7533}&\textbf{0.8608}& +14.27\%\\
\cmidrule{2-16}&NDCG@5 &0.0948& 0.1676 &0.2129& 0.1813& 0.1695 & \underline{0.2444}& 0.2428&0.2158& 0.1973 &0.2297 &0.2138&0.2171&\textbf{0.2632}& +7.69\%\\
&NDCG@10 &0.1269& 0.2033 &0.2509& 0.2194& 0.2065 & 0.2835& 0.2825&0.2555& 0.2376 &0.2697 &0.2432&\underline{0.2899}&\textbf{0.3379}& +16.56\%\\
&NDCG@20 &0.1641& 0.2403 &0.2869& 0.2581& 0.2446 & 0.3181& 0.3182&0.2924& 0.2767 &0.3068 &0.2787&\underline{0.3489}&\textbf{0.3914}& +12.18\%\\
\bottomrule
\end{tabular}}
}
\end{table*}

\section{Experiments}
We conducted experiments to answer the following RQs: 

\begin{itemize}[leftmargin=*]
    \item \textbf{RQ1}: How does \method perform compared to current baselines?
    \item \textbf{RQ2}: How much of a performance enhancement do various components of \method provide? 
    \item \textbf{RQ3}: How do the various hyperparameters of \method affect the performance?
    \item \textbf{RQ4}: Whether the semantic IDs learned by \method effectively model the pattern of representations?
\end{itemize}

\begin{table}[t]
    \caption{Overview of the datasets.}
    
    \label{tab:dataset}
    \begin{tabular}{c|cccc}
    \toprule
    \textbf{Dataset} & \textbf{\#Users} & \textbf{\#Items} & \textbf{\#Inter.}&\textbf{Density}\\ \midrule
    \textbf{MovieLens}& 6,034& 3,113& 1,000,209&5.32\%\\
    \textbf{Book}& 55,907& 54,064& 1,336,796&0.44\textperthousand\\
    \textbf{Netflix}& 5,808& 8,391& 1,447,940&2.97\%\\\bottomrule
    \end{tabular}%
\end{table}

\subsection{Experimental Settings}

\subsubsection{Datasets}
We selected three public datasets that contain timestamps to model recent historical sequence, including \textbf{MovieLens}, \textbf{Book}, and \textbf{Netflix}. For the label processing of the datasets, the user ratings lie between 0 and 5, while we consider the interaction data with ratings above 3 as positive samples. We split the dataset into training, validation, and test sets in a ratio of 8:1:1. The information on datasets is conducted in Table~\ref{tab:dataset}.

\textbf{MovieLens}~\cite{MovieLens} is a dataset of user ratings for movies. We selected MovieLens-1M datasets, containing 1,000,209 interactions between 6,034 users and 3,113 movies.

\textbf{Book}~\cite{Book} is a part of the \textbf{Amazon-2023} dataset, which collects book reviews from purchasers on the Amazon e-shopping site. We retained users with 30 and items with 10 or more interactions. After filtering out the less-interacted users and items, we remain 55,907 users and 54,064 books with 1,336,796 interactions.

\textbf{Netflix}~\cite{LLMRec} is a dataset of user ratings for movies that are sourced from the Netflix Prize Data, containing 1,447,940 interactions between 5,808 users and 8,391 movies.

\subsubsection{Baselines}
We compare our proposed \method with the following 12 baselines.

- \textbf{NCF}~\cite{NCF} is one of the early methods that introduced deep learning into recommendation systems, leveraging MLP to model the similarity between users and items. 

- \textbf{NGCF}~\cite{NGCF} introduces GNN to aggregate information from neighboring nodes, generating embeddings by leveraging high-order connection signals. 

- \textbf{Bert4Rec}~\cite{Bert4Rec} utilizes a bi-directional Transformer
as the backbone to model the historical sequence of users. 

- \textbf{LightGCN}~\cite{LightGCN} simplifies the network by removing the nonlinear projections and embedding transformations present in NGCF and optimizes performance. 

- \textbf{HCCF}~\cite{HCCF} employs a hypergraph GNN to generate self-super-vised signals, which are utilized for contrastive learning. 

- \textbf{DuoRec}~\cite{DuoRec} improves modeling performance through model-level expansion and contrastive normalization. 

- \textbf{CL4SRec}~\cite{CL4SRec} generates more consistent representations by reordering and masking the sequence, followed by contrastive learning with the original sequence.

- \textbf{ICLRec}~\cite{ICRec} optimizes recommendation performance by generating clusters that represent the different user interests.

- \textbf{SimRec}~\cite{SimRec} improves performance by introducing an MLP network and performing knowledge distillation from a pre-trained GNN network for denoising and enhancement. 

- \textbf{MAERec}~\cite{MAERec} designs a task-adaptive augmentation mechanism based on GNN to enhance representation.

- \textbf{LLMRec}~\cite{LLMRec} expands the potential edges in GNNs using LLMs and generates optimized user and item profiles. 

- \textbf{VQRec}~\cite{VQ-Rec} generates semantic IDs for items using VQ-VAE through text and incorporates them into recommendations.

\subsubsection{Evaluation Metrics}
In our setup, we utilize the widely adopted Recall@K and NDCG@K~\cite{NDCG} metrics to evaluate the performance of various recommendation models. Since ranking the user's scores for all items is time-consuming, for a user, we take the interacted item as the positive sample and all the non-interacted items in the current batch as negative samples and compute the above metrics. Finally, we report the average values of the metrics across all samples in the test set.

\subsubsection{Hyperparameters}
During the training process, we utilize Adam~\cite{Adam} as the optimizer for gradient descent with $\alpha$ set to 1e-3. The batch size is set to 1024, and the embedding dimension is 64 for all types.
For the \VAE codebook, the commitment weight $\beta$ is set to 0.25, the layer of codebooks $L$ is 4, with the size of each embedding codebook $J$ set to 128. In the relation augmentation, we selected $K=30$ neighbors for augmentation; in the user latent interest exploration, we additionally selected $K'=2$ users for each dimension. The parameters of the baselines remain the same.

\subsection{Overall Performance (RQ1)}

Tab.~\ref{tab:perf} presents the top-K evaluation results of various methods on the dataset, where K is set to 5, 10, and 20. Based on the experimental results, we make the following observations:

Our proposed \method outperforms baseline methods across various datasets, demonstrating its effectiveness and highlighting the improvements brought by the augmentation of homogeneous linkages and feature enhancement within the system. Notably, we observe that while \method achieves significant improvements in Recall, the relative increase in NDCG is minor. This suggests that the model exhibits a stronger capability for item retrieval, aligning closely with the requirements of real-world recommender systems.

Firstly, NCF, which relies solely on embeddings for recommendation, demonstrates relatively poor performance. NCF does not leverage information from neighbors, highlighting the necessity of modeling interaction sequences for improved performance.

We observe that GNN-based CF methods perform relatively worse on the MovieLens and Netflix datasets compared to the Books dataset. Experimental analysis reveals that the Books dataset features a large number of users and items, resulting in very low data sparsity. Consequently, the average number of interacted items per user is relatively low, making it more challenging to capture users’ dynamic interests. GNN-based methods, which aggregate information from all neighbors indiscriminately, perform better on the Books dataset but show inferior performance on datasets where users interact with more items on average.

Similarly, the performance improvement of our \method on the Books dataset is relatively minor. This is also due to the shorter average interaction sequences of users and items, which limit the generalization and information diffusion effects achieved through dynamic feature enhancement with semantic IDs.

LLMRec, which enhances representations using LLM, demonstrates superior performance. This further validates the effectiveness of enhancing user and item representations. However, LLMRec is still constrained by the limitations of GNN-based methods, as it cannot effectively model the real-time features of users and items. Additionally, although LLMRec addresses the sparsity of interaction data by enhancing item-user connections, it overlooks the importance of reinforcing homogeneous connections within users and within items, therefore fails to achieve optimal performance.

For sequence-based recommendation algorithms, various methods exhibit relatively consistent performance across datasets.
Notably, ICLRec clusters user interests, which share a similar concept with the proposed \method. However, its relatively poor performance on multiple datasets may be attributed to its single clustering approach, which inadequately explores user interests, leading to suboptimal user representation modeling.
MAERec leverages GNN to model user sequences and extract representations but neglects the exploration of user-user homogeneous relationships and feature enhancement, which contributes to its weaker performance. DuoRec performs relatively well among the baselines, surpassing some of the more recent approaches. This is because DuoRec’s design also measures the semantics of sequences based on user interaction sequences, and assists in training by leveraging the semantic similarity of different sequences. This demonstrates the effectiveness of extracting semantic patterns from sequences. However, its neglect of joint attribute modeling results in our \method outperforming it.

VQRec utilizes VQ-VAE for item modeling and performs sequential recommendations based on the generated semantic IDs. The metrics reflect the superiority of semantic ID modeling compared to other baselines. However, it is still limited by coarse-grained attributes and does not model user behavior. This neglect of user patterns embedded in the sequences leads to suboptimal results.

\subsection{Ablation Study (RQ2)}
In this subsection, we conduct ablation studies on the proposed \method to analyze the effectiveness of its components. The experimental results are presented in Tab.~\ref{tab:ablation}.

\begin{table}[t]
    \centering
    \caption{Ablation study of key components of \method.}
    {\small
    \begin{tabular}{c|c|cc|cc|cc}
        \toprule
        \multicolumn{2}{c|}{Dataset} & \multicolumn{2}{c|}{\textbf{MovieLens}} & \multicolumn{2}{c|}{\textbf{Books}} & \multicolumn{2}{c}{\textbf{Netflix}} \\
        \midrule \multicolumn{2}{c|}{Metric} & R@5 & N@5 & R@5 & N@5& R@5 & N@5 \\ 
        \midrule
        \multirow{3}{*}{\makecell{\texttt{w/o} \\ Feat. \\ Aug.}}& User  & 0.5515& 0.3533& 0.3575& 0.2121& 0.3846& 0.2392\\
        & Item & 0.5709& 0.3698& 0.3913& 0.2363& 0.4093& 0.2564\\
        & All & 0.5495& 0.3508& 0.3759& 0.2256& 0.3736&0.2319\\
        \midrule
        \multirow{4}{*}{\makecell{\texttt{w/o} \\ Relat. \\ Aug.}}
        & Lat.& 0.6066& 0.3954& 0.4210& 0.2572& 0.4099& 0.2570\\
        & User& 0.5945& 0.3863& 0.3742& 0.2232& 0.4062& 0.2540\\
        & Item& 0.5448& 0.3465& 0.3830& 0.2303& 0.4033& 0.2515\\
        & All& 0.5816& 0.3750& 0.3829& 0.2296& 0.3855& 0.2411\\
        \midrule
        \multicolumn{2}{c|}{\method} & 0.6204& 0.4051& 0.4261& 0.2608& 0.4190& 0.2632\\
        \bottomrule
    \end{tabular}
    \label{tab:ablation}
    }
\end{table}

\begin{figure*}[t]
    \centering
    \subfigure[Impact of similar user\&item numbers]{
        \includegraphics[width=.32\linewidth]{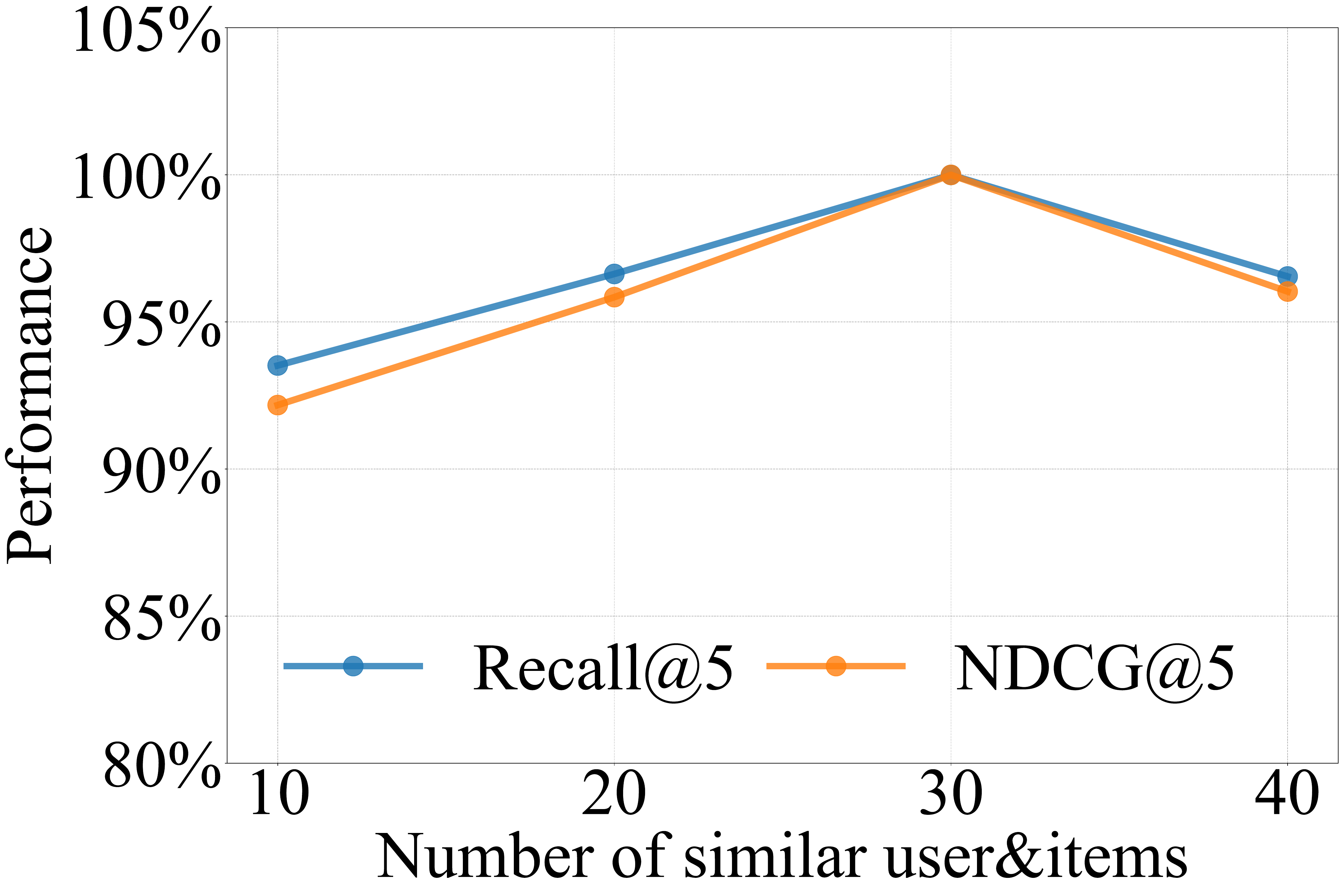}
        \label{fig:sim}
    }
    \subfigure[Impact of codebook embedding size]{
        \includegraphics[width=.32\linewidth]{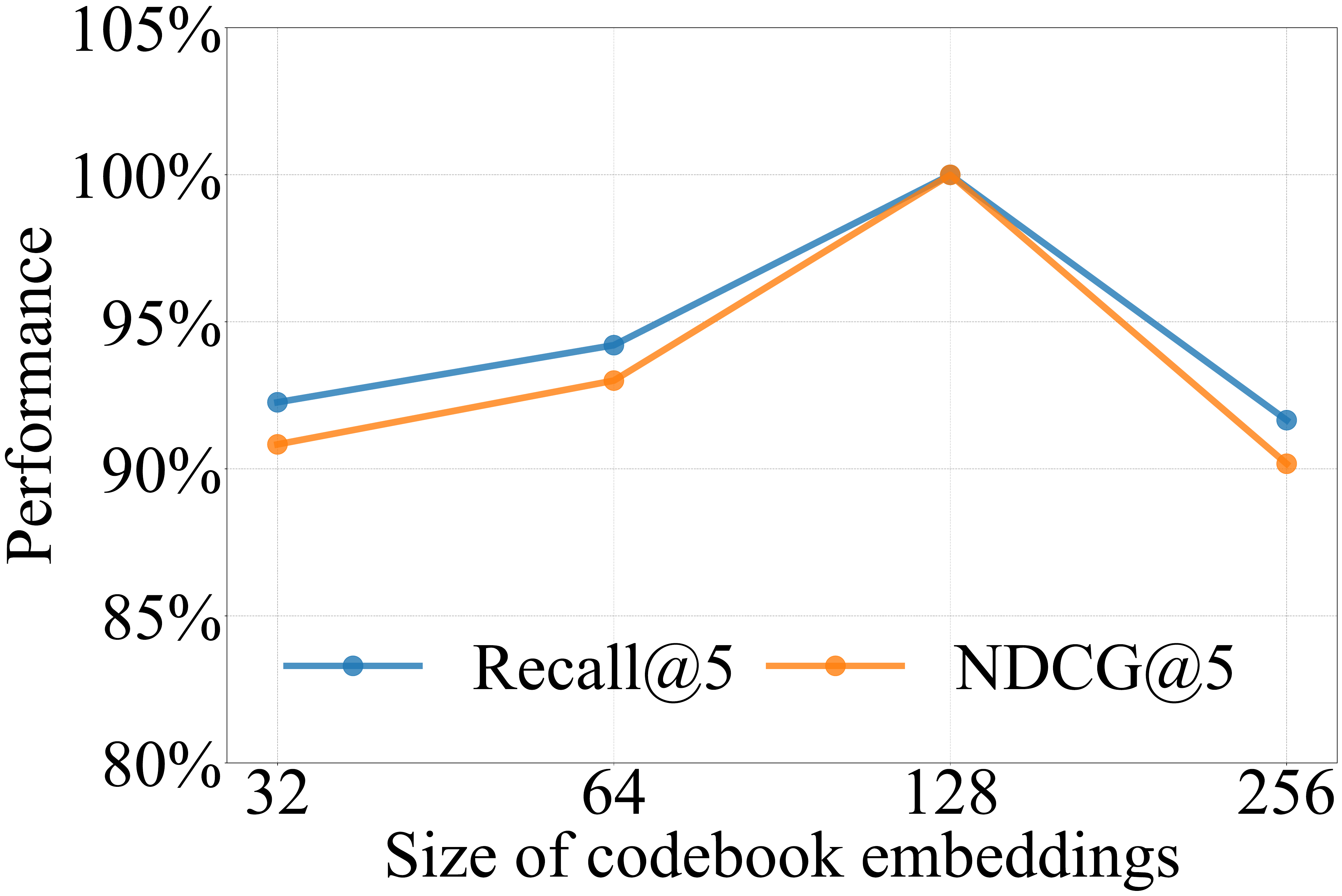}
        \label{fig:code}
    }
    \subfigure[Impact of codebook layers]{
        \includegraphics[width=.32\linewidth]{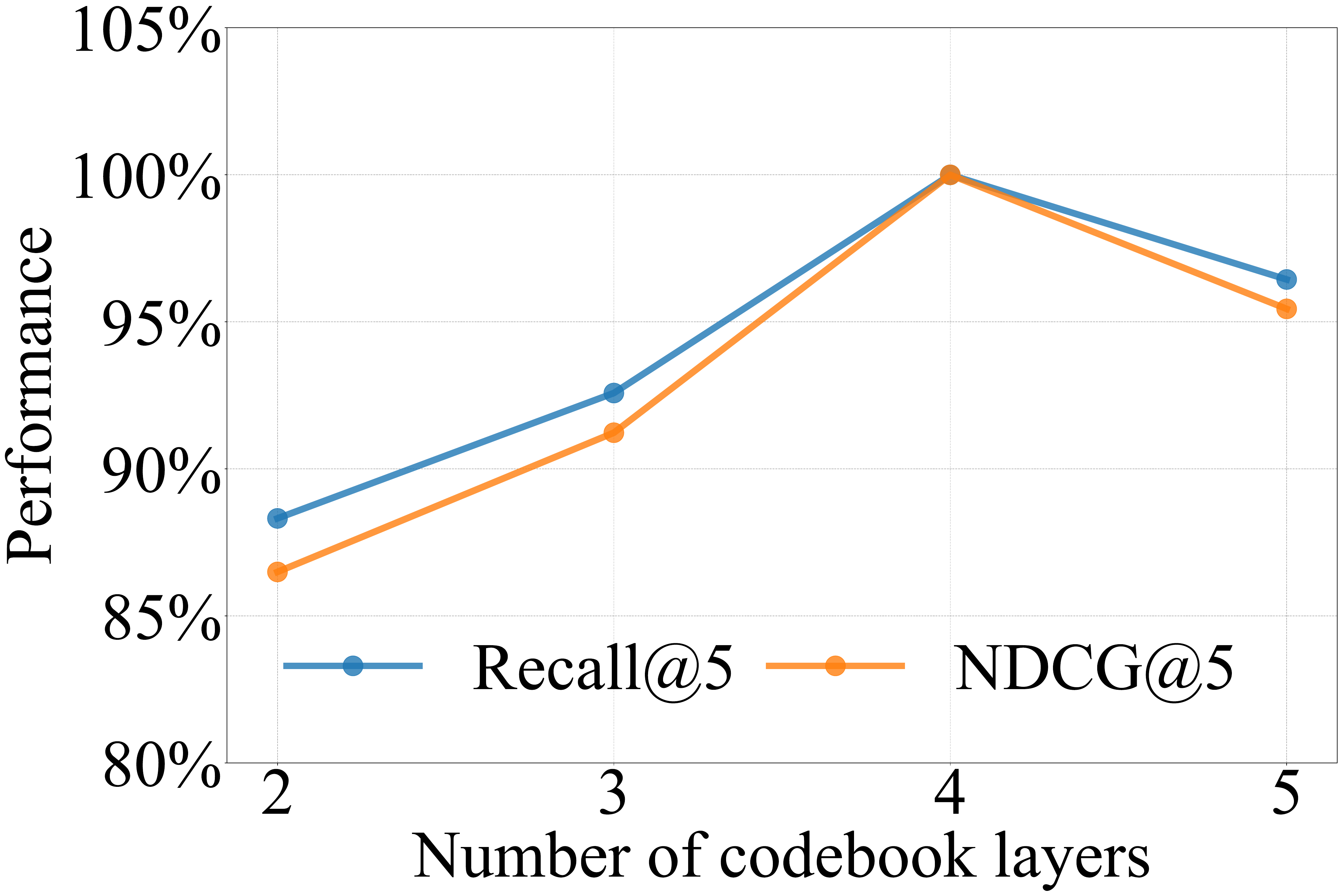}
        \label{fig:layer}
    }
    \caption{Performance of \method on MovieLens dataset \textit{w.r.t.} different hyperparameters. For both the number of neighbors selected and the size of the codebook, values that are too low or too high can lead to a decline in recommendation performance.}
    \label{fig:param}
\end{figure*}

\subsubsection{Effect of Feature Augmentation.}
Feature augmentation leverages the semantic IDs generated from the dynamic representations of users and items as additional features to optimize recommendation performance. By removing embeddings from different parts $\boldsymbol{e}_u^s$, $\boldsymbol{e}_i^s$ from the input embedding in Eq.~(\ref{equ:input}), we analyze the improvements brought by feature augmentation for both users and items. First, removing any type of feature augmentation results in worse performance, highlighting the effectiveness of our feature augmentation approach. We observe that removing the augmented user features leads to a more significant performance degradation compared to removing the item features. This indicates the model’s enhanced ability to capture user interests more effectively through semantic ID feature augmentation. Moreover, we observe that aligning user and item representations in the dynamic representation generator can lead to performance degradation on certain datasets (Books) if representation augmentation is applied to only items.

\subsubsection{Effect of Linkage Augmentation.}
Lat. represents the impact of removing neighboring selected through potential pattern exploration $\mathcal{Q}^{\mathrm{lat}, l}$ from $\mathcal{Q}^{\mathrm{all}}$, retaining only those selected based on explicit patterns.
We observe that removing neighbors obtained through exploration results in slightly poorer performance, which demonstrates the effectiveness of exploring potential patterns.

Similarly, as above section, we remove the corresponding embeddings from the input embeddings to conduct an ablation study on linkage augmentation. It can be observed that removing the augmentation embeddings for either users or items results in a decline in performance.
Meanwhile, for the Books dataset, the effect of user linkage augmentation is more significant compared with other datasets, due to the larger number of users in the dataset.

\subsection{Parameter Analysis (RQ3)}
We analyzed the parameter sensitivity of \method in three dimensions: numbers of similar users\&items $K$, size of codebook $J$, and the number of codebook layers $L$. Results are displayed in Fig.~\ref{fig:param}.

\subsubsection{Impact of Similar User\&Item Numbers in Linkage Augmentation}

We evaluated the impact of the number of neighboring users/items selected during linkage expansion on performance, shown in Fig.~\ref{fig:sim}. It can be observed that when too few similar neighbors are selected, the system’s information diffusion efficiency is low, resulting in degraded performance. Conversely, selecting too many neighbors may include users/items with lower similarity. The system becomes overly generalized, lacking accurate personalized information propagation, leading to the incorrect diffusion of collaborative information, which also results in inferior outcomes.

\subsubsection{Impact of Codebook Embedding Size in \VAE}

Regarding the impact of different codebook sizes on performance in Fig.~\ref{fig:code}, we observe that when the codebook size is small, the system cannot effectively quantify the representations of users\&items, leading to insufficient differentiation between users and an inability to capture user features. On the other hand, when the size is large, the quantization effect of the VAE deteriorates, resulting in poor generalization performance. This leads to fewer shared semantic IDs between users\&items, reducing information diffusion efficiency.

\subsubsection{Impact of Codebook Layer Numbers in \VAE}

The number of codebook layers in the VAE also has a significant impact on performance, as shown in Fig.~\ref{fig:layer}. When the number of codebook layers decreases, the model’s performance drops substantially, highlighting the necessity of modeling multi-dimensional dynamic representations. However, when the number of codebook layers becomes too large, a similar issue of degraded generalization performance arises, as discussed in the previous section, leading to a decline in the model’s effectiveness. This differs from previous vector quantization approaches based on RQ-VAE~\cite{TIGER}. In semantic ID generation with RQ-VAE, a greater number of codebook layers implies a finer quantization granularity, which often does not result in performance degradation. In contrast, \VAE, benefiting from its information-decoupled decomposition, does not share the same requirements as RQ-VAE and offers higher interpretability.

\subsection{Case Study (RQ4)}

\begin{figure}[t]
    \centering
    \includegraphics[width=\linewidth]{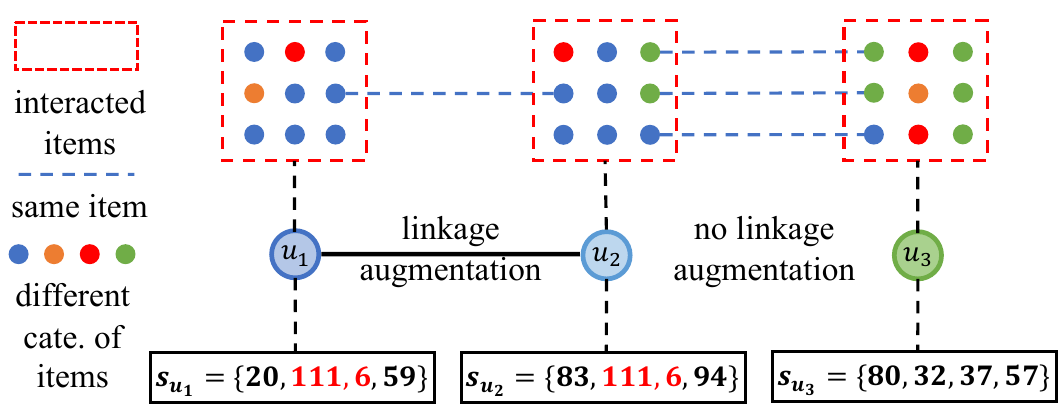}
    \caption{Categories of items in interaction sequence of users, and the semantic ID patterns generated. \method can identify the similarity of interest patterns between users without being affected by overlapping interaction sequences.}
    \label{fig:case}
\end{figure}

We qualitatively investigate the effects of our proposed \VAE can jointly extract behavioral patterns from interaction sequences and attributes, thereby enabling more accurate homogeneous connection augmentation.
To illustrate this, we selected a pair of users in the MovieLens dataset and analyzed their interaction sequences, the categories of interacted items, and the semantic ID patterns generated through \VAE, as shown in Fig.~\ref{fig:case}. 

It can be observed that for the first two users, their interaction sequences share only one overlapping item, which would typically be overlooked using traditional two-hop neighbor methods for linkage. However, most of their interacted items belong to the same category, indicating similar behavioral patterns. Our \VAE effectively extracts patterns from the sequences and explicitly shows their similarity (partial overlap), thereby enabling linkage augmentation.
In contrast, the last two users have more overlapping interaction sequences, yet their interests are quite different based on the sequence data. 
Traditional augmentation approaches, due to the high overlap of their interaction sequences, may incorrectly identify users as having similar patterns, thereby leading to erroneous linkage augmentation.
Our \VAE captures this information and avoids making connections based on unrelated semantic ID features. This demonstrates the efficiency of our designed \VAE in jointly extracting patterns from sequences and attributes.

\section{Conclusion}

In this paper, we propose a novel two-stage collaborative filtering algorithm \method based on VQ.
\method first employs the newly proposed \VAE to efficiently quantize and extract multidimensional, decoupled semantic ID patterns jointly from interaction and attributes for users and items.
The generated pattern sequences are then utilized for collaborative filtering augmentation. This approach augments sparse homogeneous linkages and insufficient user and item attributes in the system, thereby improving the efficiency of information diffusion in collaborative filtering. Comprehensive experiments demonstrate the effectiveness of our algorithm, showing its capability to effectively model behavior patterns, and optimize recommendation abilities.

\newpage
\bibliographystyle{ACM-Reference-Format}
\bibliography{reference}


\begin{thebibliography}{48}


\ifx \showCODEN    \undefined \def \showCODEN     #1{\unskip}     \fi
\ifx \showDOI      \undefined \def \showDOI       #1{#1}\fi
\ifx \showISBNx    \undefined \def \showISBNx     #1{\unskip}     \fi
\ifx \showISBNxiii \undefined \def \showISBNxiii  #1{\unskip}     \fi
\ifx \showISSN     \undefined \def \showISSN      #1{\unskip}     \fi
\ifx \showLCCN     \undefined \def \showLCCN      #1{\unskip}     \fi
\ifx \shownote     \undefined \def \shownote      #1{#1}          \fi
\ifx \showarticletitle \undefined \def \showarticletitle #1{#1}   \fi
\ifx \showURL      \undefined \def \showURL       {\relax}        \fi
\providecommand\bibfield[2]{#2}
\providecommand\bibinfo[2]{#2}
\providecommand\natexlab[1]{#1}
\providecommand\showeprint[2][]{arXiv:#2}

\bibitem[Agfarap(2019)]%
        {ReLU}
\bibfield{author}{\bibinfo{person}{Abien~Fred Agfarap}.} \bibinfo{year}{2019}\natexlab{}.
\newblock \bibinfo{title}{Deep Learning using Rectified Linear Units (ReLU)}.
\newblock
\newblock
\showeprint[arxiv]{1803.08375}~[cs.NE]


\bibitem[Balen and Levy(2019)]%
        {PQVAE}
\bibfield{author}{\bibinfo{person}{Jan~Van Balen} {and} \bibinfo{person}{Mark Levy}.} \bibinfo{year}{2019}\natexlab{}.
\newblock \showarticletitle{PQ-VAE: Efficient Recommendation Using Quantized Embeddings}. In \bibinfo{booktitle}{\emph{ACM Conference on Recommender Systems}}.
\newblock
\urldef\tempurl%
\url{https://api.semanticscholar.org/CorpusID:201672459}
\showURL{%
\tempurl}


\bibitem[Cao et~al\mbox{.}(2022)]%
        {GIFT}
\bibfield{author}{\bibinfo{person}{Yi Cao}, \bibinfo{person}{Sihao Hu}, \bibinfo{person}{Yu Gong}, \bibinfo{person}{Zhao Li}, \bibinfo{person}{Yazheng Yang}, \bibinfo{person}{Qingwen Liu}, {and} \bibinfo{person}{Shouling Ji}.} \bibinfo{year}{2022}\natexlab{}.
\newblock \showarticletitle{GIFT: Graph-guIded Feature Transfer for Cold-Start Video Click-Through Rate Prediction}. In \bibinfo{booktitle}{\emph{Proceedings of the 31st ACM International Conference on Information \& Knowledge Management}} (Atlanta, GA, USA) \emph{(\bibinfo{series}{CIKM '22})}. \bibinfo{publisher}{Association for Computing Machinery}, \bibinfo{address}{New York, NY, USA}, \bibinfo{pages}{2964–2973}.
\newblock
\showISBNx{9781450392365}
\urldef\tempurl%
\url{https://doi.org/10.1145/3511808.3557120}
\showDOI{\tempurl}


\bibitem[Chen et~al\mbox{.}(2024)]%
        {M3CSR}
\bibfield{author}{\bibinfo{person}{Gaode Chen}, \bibinfo{person}{Ruina Sun}, \bibinfo{person}{Yuezihan Jiang}, \bibinfo{person}{Jiangxia Cao}, \bibinfo{person}{Qi Zhang}, \bibinfo{person}{Jingjian Lin}, \bibinfo{person}{Han Li}, \bibinfo{person}{Kun Gai}, {and} \bibinfo{person}{Xinghua Zhang}.} \bibinfo{year}{2024}\natexlab{}.
\newblock \showarticletitle{A Multi-modal Modeling Framework for Cold-start Short-video Recommendation}. In \bibinfo{booktitle}{\emph{Proceedings of the 18th ACM Conference on Recommender Systems}}. \bibinfo{pages}{391--400}.
\newblock


\bibitem[Chen et~al\mbox{.}(2022)]%
        {ICRec}
\bibfield{author}{\bibinfo{person}{Yongjun Chen}, \bibinfo{person}{Zhiwei Liu}, \bibinfo{person}{Jia Li}, \bibinfo{person}{Julian McAuley}, {and} \bibinfo{person}{Caiming Xiong}.} \bibinfo{year}{2022}\natexlab{}.
\newblock \showarticletitle{Intent Contrastive Learning for Sequential Recommendation}. In \bibinfo{booktitle}{\emph{Proceedings of the ACM Web Conference 2022}} (Virtual Event, Lyon, France) \emph{(\bibinfo{series}{WWW '22})}. \bibinfo{publisher}{Association for Computing Machinery}, \bibinfo{address}{New York, NY, USA}, \bibinfo{pages}{2172–2182}.
\newblock
\showISBNx{9781450390965}
\urldef\tempurl%
\url{https://doi.org/10.1145/3485447.3512090}
\showDOI{\tempurl}


\bibitem[Devlin et~al\mbox{.}(2019)]%
        {BERT}
\bibfield{author}{\bibinfo{person}{Jacob Devlin}, \bibinfo{person}{Ming-Wei Chang}, \bibinfo{person}{Kenton Lee}, {and} \bibinfo{person}{Kristina Toutanova}.} \bibinfo{year}{2019}\natexlab{}.
\newblock \bibinfo{title}{BERT: Pre-training of Deep Bidirectional Transformers for Language Understanding}.
\newblock
\newblock
\showeprint[arxiv]{1810.04805}~[cs.CL]
\urldef\tempurl%
\url{https://arxiv.org/abs/1810.04805}
\showURL{%
\tempurl}


\bibitem[Harper and Konstan(2015)]%
        {MovieLens}
\bibfield{author}{\bibinfo{person}{F.~Maxwell Harper} {and} \bibinfo{person}{Joseph~A. Konstan}.} \bibinfo{year}{2015}\natexlab{}.
\newblock \showarticletitle{The MovieLens Datasets: History and Context}.
\newblock \bibinfo{journal}{\emph{ACM Trans. Interact. Intell. Syst.}} \bibinfo{volume}{5}, \bibinfo{number}{4}, Article \bibinfo{articleno}{19} (\bibinfo{date}{dec} \bibinfo{year}{2015}), \bibinfo{numpages}{19}~pages.
\newblock
\showISSN{2160-6455}
\urldef\tempurl%
\url{https://doi.org/10.1145/2827872}
\showDOI{\tempurl}


\bibitem[He et~al\mbox{.}(2020)]%
        {LightGCN}
\bibfield{author}{\bibinfo{person}{Xiangnan He}, \bibinfo{person}{Kuan Deng}, \bibinfo{person}{Xiang Wang}, \bibinfo{person}{Yan Li}, \bibinfo{person}{YongDong Zhang}, {and} \bibinfo{person}{Meng Wang}.} \bibinfo{year}{2020}\natexlab{}.
\newblock \showarticletitle{LightGCN: Simplifying and Powering Graph Convolution Network for Recommendation}. In \bibinfo{booktitle}{\emph{Proceedings of the 43rd International ACM SIGIR Conference on Research and Development in Information Retrieval}} (Virtual Event, China) \emph{(\bibinfo{series}{SIGIR '20})}. \bibinfo{publisher}{Association for Computing Machinery}, \bibinfo{address}{New York, NY, USA}, \bibinfo{pages}{639–648}.
\newblock
\showISBNx{9781450380164}
\urldef\tempurl%
\url{https://doi.org/10.1145/3397271.3401063}
\showDOI{\tempurl}


\bibitem[He et~al\mbox{.}(2017)]%
        {NCF}
\bibfield{author}{\bibinfo{person}{Xiangnan He}, \bibinfo{person}{Lizi Liao}, \bibinfo{person}{Hanwang Zhang}, \bibinfo{person}{Liqiang Nie}, \bibinfo{person}{Xia Hu}, {and} \bibinfo{person}{Tat-Seng Chua}.} \bibinfo{year}{2017}\natexlab{}.
\newblock \showarticletitle{Neural Collaborative Filtering}. In \bibinfo{booktitle}{\emph{Proceedings of the 26th International Conference on World Wide Web}} (Perth, Australia) \emph{(\bibinfo{series}{WWW '17})}. \bibinfo{publisher}{International World Wide Web Conferences Steering Committee}, \bibinfo{address}{Republic and Canton of Geneva, CHE}, \bibinfo{pages}{173–182}.
\newblock
\showISBNx{9781450349130}
\urldef\tempurl%
\url{https://doi.org/10.1145/3038912.3052569}
\showDOI{\tempurl}


\bibitem[Hendrycks and Gimpel(2023)]%
        {GELU}
\bibfield{author}{\bibinfo{person}{Dan Hendrycks} {and} \bibinfo{person}{Kevin Gimpel}.} \bibinfo{year}{2023}\natexlab{}.
\newblock \bibinfo{title}{Gaussian Error Linear Units (GELUs)}.
\newblock
\newblock
\showeprint[arxiv]{1606.08415}~[cs.LG]
\urldef\tempurl%
\url{https://arxiv.org/abs/1606.08415}
\showURL{%
\tempurl}


\bibitem[Hidasi et~al\mbox{.}(2016)]%
        {GRU4Rec}
\bibfield{author}{\bibinfo{person}{Balázs Hidasi}, \bibinfo{person}{Alexandros Karatzoglou}, \bibinfo{person}{Linas Baltrunas}, {and} \bibinfo{person}{Domonkos Tikk}.} \bibinfo{year}{2016}\natexlab{}.
\newblock \bibinfo{title}{Session-based Recommendations with Recurrent Neural Networks}.
\newblock
\newblock
\showeprint[arxiv]{1511.06939}~[cs.LG]
\urldef\tempurl%
\url{https://arxiv.org/abs/1511.06939}
\showURL{%
\tempurl}


\bibitem[Hou et~al\mbox{.}(2023)]%
        {VQ-Rec}
\bibfield{author}{\bibinfo{person}{Yupeng Hou}, \bibinfo{person}{Zhankui He}, \bibinfo{person}{Julian McAuley}, {and} \bibinfo{person}{Wayne~Xin Zhao}.} \bibinfo{year}{2023}\natexlab{}.
\newblock \showarticletitle{Learning vector-quantized item representation for transferable sequential recommenders}. In \bibinfo{booktitle}{\emph{Proceedings of the ACM Web Conference 2023}}. \bibinfo{pages}{1162--1171}.
\newblock


\bibitem[Hou et~al\mbox{.}(2024)]%
        {Book}
\bibfield{author}{\bibinfo{person}{Yupeng Hou}, \bibinfo{person}{Jiacheng Li}, \bibinfo{person}{Zhankui He}, \bibinfo{person}{An Yan}, \bibinfo{person}{Xiusi Chen}, {and} \bibinfo{person}{Julian McAuley}.} \bibinfo{year}{2024}\natexlab{}.
\newblock \showarticletitle{Bridging Language and Items for Retrieval and Recommendation}.
\newblock \bibinfo{journal}{\emph{arXiv preprint arXiv:2403.03952}} (\bibinfo{year}{2024}).
\newblock


\bibitem[Hu et~al\mbox{.}(2024)]%
        {user-side-info}
\bibfield{author}{\bibinfo{person}{Zheng Hu}, \bibinfo{person}{Zhe Li}, \bibinfo{person}{Ziyun Jiao}, \bibinfo{person}{Satoshi Nakagawa}, \bibinfo{person}{Jiawen Deng}, \bibinfo{person}{Shimin Cai}, \bibinfo{person}{Tao Zhou}, {and} \bibinfo{person}{Fuji Ren}.} \bibinfo{year}{2024}\natexlab{}.
\newblock \bibinfo{title}{Bridging the User-side Knowledge Gap in Knowledge-aware Recommendations with Large Language Models}.
\newblock
\newblock
\showeprint[arxiv]{2412.13544}~[cs.IR]
\urldef\tempurl%
\url{https://arxiv.org/abs/2412.13544}
\showURL{%
\tempurl}


\bibitem[Huang et~al\mbox{.}(2013)]%
        {dual-tower}
\bibfield{author}{\bibinfo{person}{Po-Sen Huang}, \bibinfo{person}{Xiaodong He}, \bibinfo{person}{Jianfeng Gao}, \bibinfo{person}{Li Deng}, \bibinfo{person}{Alex Acero}, {and} \bibinfo{person}{Larry Heck}.} \bibinfo{year}{2013}\natexlab{}.
\newblock \showarticletitle{Learning deep structured semantic models for web search using clickthrough data}. In \bibinfo{booktitle}{\emph{Proceedings of the 22nd ACM international conference on Information \& Knowledge Management}}. \bibinfo{pages}{2333--2338}.
\newblock


\bibitem[Jiang et~al\mbox{.}(2024)]%
        {SHaRe}
\bibfield{author}{\bibinfo{person}{Wei Jiang}, \bibinfo{person}{Xinyi Gao}, \bibinfo{person}{Guandong Xu}, \bibinfo{person}{Tong Chen}, {and} \bibinfo{person}{Hongzhi Yin}.} \bibinfo{year}{2024}\natexlab{}.
\newblock \bibinfo{title}{Challenging Low Homophily in Social Recommendation}.
\newblock
\newblock
\showeprint[arxiv]{2401.14606}~[cs.IR]
\urldef\tempurl%
\url{https://arxiv.org/abs/2401.14606}
\showURL{%
\tempurl}


\bibitem[Kang and McAuley(2018)]%
        {SASRec}
\bibfield{author}{\bibinfo{person}{Wang-Cheng Kang} {and} \bibinfo{person}{Julian McAuley}.} \bibinfo{year}{2018}\natexlab{}.
\newblock \bibinfo{title}{Self-Attentive Sequential Recommendation}.
\newblock
\newblock
\showeprint[arxiv]{1808.09781}~[cs.IR]
\urldef\tempurl%
\url{https://arxiv.org/abs/1808.09781}
\showURL{%
\tempurl}


\bibitem[Kim et~al\mbox{.}(2024a)]%
        {A-LLM}
\bibfield{author}{\bibinfo{person}{Sein Kim}, \bibinfo{person}{Hongseok Kang}, \bibinfo{person}{Seungyoon Choi}, \bibinfo{person}{Donghyun Kim}, \bibinfo{person}{Minchul Yang}, {and} \bibinfo{person}{Chanyoung Park}.} \bibinfo{year}{2024}\natexlab{a}.
\newblock \showarticletitle{Large Language Models meet Collaborative Filtering: An Efficient All-round LLM-based Recommender System}. In \bibinfo{booktitle}{\emph{Proceedings of the 30th ACM SIGKDD Conference on Knowledge Discovery and Data Mining}} (Barcelona, Spain) \emph{(\bibinfo{series}{KDD '24})}. \bibinfo{publisher}{Association for Computing Machinery}, \bibinfo{address}{New York, NY, USA}, \bibinfo{pages}{1395–1406}.
\newblock
\showISBNx{9798400704901}
\urldef\tempurl%
\url{https://doi.org/10.1145/3637528.3671931}
\showDOI{\tempurl}


\bibitem[Kim et~al\mbox{.}(2024b)]%
        {LLMRec}
\bibfield{author}{\bibinfo{person}{Sein Kim}, \bibinfo{person}{Hongseok Kang}, \bibinfo{person}{Seungyoon Choi}, \bibinfo{person}{Donghyun Kim}, \bibinfo{person}{Minchul Yang}, {and} \bibinfo{person}{Chanyoung Park}.} \bibinfo{year}{2024}\natexlab{b}.
\newblock \showarticletitle{Large Language Models meet Collaborative Filtering: An Efficient All-round LLM-based Recommender System}. In \bibinfo{booktitle}{\emph{Proceedings of the 30th ACM SIGKDD Conference on Knowledge Discovery and Data Mining}} (Barcelona, Spain) \emph{(\bibinfo{series}{KDD '24})}. \bibinfo{publisher}{Association for Computing Machinery}, \bibinfo{address}{New York, NY, USA}, \bibinfo{pages}{1395–1406}.
\newblock
\showISBNx{9798400704901}
\urldef\tempurl%
\url{https://doi.org/10.1145/3637528.3671931}
\showDOI{\tempurl}


\bibitem[Kingma and Ba(2014)]%
        {Adam}
\bibfield{author}{\bibinfo{person}{Diederik~P. Kingma} {and} \bibinfo{person}{Jimmy Ba}.} \bibinfo{year}{2014}\natexlab{}.
\newblock \bibinfo{title}{Adam: A Method for Stochastic Optimization}.
\newblock
\newblock
\urldef\tempurl%
\url{https://doi.org/10.48550/ARXIV.1412.6980}
\showDOI{\tempurl}


\bibitem[Klema and Laub(1980)]%
        {SVD}
\bibfield{author}{\bibinfo{person}{V. Klema} {and} \bibinfo{person}{A. Laub}.} \bibinfo{year}{1980}\natexlab{}.
\newblock \showarticletitle{The singular value decomposition: Its computation and some applications}.
\newblock \bibinfo{journal}{\emph{IEEE Trans. Automat. Control}} \bibinfo{volume}{25}, \bibinfo{number}{2} (\bibinfo{year}{1980}), \bibinfo{pages}{164--176}.
\newblock
\urldef\tempurl%
\url{https://doi.org/10.1109/TAC.1980.1102314}
\showDOI{\tempurl}


\bibitem[Koren et~al\mbox{.}(2009)]%
        {CF}
\bibfield{author}{\bibinfo{person}{Yehuda Koren}, \bibinfo{person}{Robert Bell}, {and} \bibinfo{person}{Chris Volinsky}.} \bibinfo{year}{2009}\natexlab{}.
\newblock \showarticletitle{Matrix Factorization Techniques for Recommender Systems}.
\newblock \bibinfo{journal}{\emph{Computer}} \bibinfo{volume}{42}, \bibinfo{number}{8} (\bibinfo{year}{2009}), \bibinfo{pages}{30--37}.
\newblock
\urldef\tempurl%
\url{https://doi.org/10.1109/MC.2009.263}
\showDOI{\tempurl}


\bibitem[Lee et~al\mbox{.}(2022)]%
        {RQ-VAE}
\bibfield{author}{\bibinfo{person}{Doyup Lee}, \bibinfo{person}{Chiheon Kim}, \bibinfo{person}{Saehoon Kim}, \bibinfo{person}{Minsu Cho}, {and} \bibinfo{person}{Wook-Shin Han}.} \bibinfo{year}{2022}\natexlab{}.
\newblock \showarticletitle{Autoregressive Image Generation using Residual Quantization}. In \bibinfo{booktitle}{\emph{2022 IEEE/CVF Conference on Computer Vision and Pattern Recognition (CVPR)}}. \bibinfo{pages}{11513--11522}.
\newblock
\urldef\tempurl%
\url{https://doi.org/10.1109/CVPR52688.2022.01123}
\showDOI{\tempurl}


\bibitem[Luo et~al\mbox{.}(2025)]%
        {PAM}
\bibfield{author}{\bibinfo{person}{Yunze Luo}, \bibinfo{person}{Yuezihan Jiang}, \bibinfo{person}{Yinjie Jiang}, \bibinfo{person}{Gaode Chen}, \bibinfo{person}{Jingchi Wang}, \bibinfo{person}{Kaigui Bian}, \bibinfo{person}{Peiyi Li}, {and} \bibinfo{person}{Qi Zhang}.} \bibinfo{year}{2025}\natexlab{}.
\newblock \showarticletitle{Online Item Cold-Start Recommendation with Popularity-Aware Meta-Learning}. In \bibinfo{booktitle}{\emph{Proceedings of the 31st ACM SIGKDD Conference on Knowledge Discovery and Data Mining V.1}} (Toronto ON, Canada) \emph{(\bibinfo{series}{KDD '25})}. \bibinfo{publisher}{Association for Computing Machinery}, \bibinfo{address}{New York, NY, USA}, \bibinfo{pages}{927–937}.
\newblock
\showISBNx{9798400712456}
\urldef\tempurl%
\url{https://doi.org/10.1145/3690624.3709336}
\showDOI{\tempurl}


\bibitem[Ma{\'c}kiewicz and Ratajczak(1993)]%
        {PCA}
\bibfield{author}{\bibinfo{person}{Andrzej Ma{\'c}kiewicz} {and} \bibinfo{person}{Waldemar Ratajczak}.} \bibinfo{year}{1993}\natexlab{}.
\newblock \showarticletitle{Principal components analysis (PCA)}.
\newblock \bibinfo{journal}{\emph{Computers \& Geosciences}} \bibinfo{volume}{19}, \bibinfo{number}{3} (\bibinfo{year}{1993}), \bibinfo{pages}{303--342}.
\newblock


\bibitem[Qiu et~al\mbox{.}(2022)]%
        {DuoRec}
\bibfield{author}{\bibinfo{person}{Ruihong Qiu}, \bibinfo{person}{Zi Huang}, \bibinfo{person}{Hongzhi Yin}, {and} \bibinfo{person}{Zijian Wang}.} \bibinfo{year}{2022}\natexlab{}.
\newblock \showarticletitle{Contrastive Learning for Representation Degeneration Problem in Sequential Recommendation}. In \bibinfo{booktitle}{\emph{Proceedings of the Fifteenth ACM International Conference on Web Search and Data Mining}} (Virtual Event, AZ, USA) \emph{(\bibinfo{series}{WSDM '22})}. \bibinfo{publisher}{Association for Computing Machinery}, \bibinfo{address}{New York, NY, USA}, \bibinfo{pages}{813–823}.
\newblock
\showISBNx{9781450391320}
\urldef\tempurl%
\url{https://doi.org/10.1145/3488560.3498433}
\showDOI{\tempurl}


\bibitem[Rajput et~al\mbox{.}(2023)]%
        {TIGER}
\bibfield{author}{\bibinfo{person}{Shashank Rajput}, \bibinfo{person}{Nikhil Mehta}, \bibinfo{person}{Anima Singh}, \bibinfo{person}{Raghunandan Hulikal~Keshavan}, \bibinfo{person}{Trung Vu}, \bibinfo{person}{Lukasz Heldt}, \bibinfo{person}{Lichan Hong}, \bibinfo{person}{Yi Tay}, \bibinfo{person}{Vinh Tran}, \bibinfo{person}{Jonah Samost}, {et~al\mbox{.}}} \bibinfo{year}{2023}\natexlab{}.
\newblock \showarticletitle{Recommender systems with generative retrieval}.
\newblock \bibinfo{journal}{\emph{Advances in Neural Information Processing Systems}}  \bibinfo{volume}{36} (\bibinfo{year}{2023}), \bibinfo{pages}{10299--10315}.
\newblock


\bibitem[Razavi et~al\mbox{.}(2019)]%
        {VQ-VAE2}
\bibfield{author}{\bibinfo{person}{Ali Razavi}, \bibinfo{person}{Aaron Van~den Oord}, {and} \bibinfo{person}{Oriol Vinyals}.} \bibinfo{year}{2019}\natexlab{}.
\newblock \showarticletitle{Generating diverse high-fidelity images with vq-vae-2}.
\newblock \bibinfo{journal}{\emph{Advances in neural information processing systems}}  \bibinfo{volume}{32} (\bibinfo{year}{2019}).
\newblock


\bibitem[Rendle et~al\mbox{.}(2009)]%
        {BPR}
\bibfield{author}{\bibinfo{person}{Steffen Rendle}, \bibinfo{person}{Christoph Freudenthaler}, \bibinfo{person}{Zeno Gantner}, {and} \bibinfo{person}{Lars Schmidt-Thieme}.} \bibinfo{year}{2009}\natexlab{}.
\newblock \showarticletitle{BPR: Bayesian personalized ranking from implicit feedback}. In \bibinfo{booktitle}{\emph{Proceedings of the Twenty-Fifth Conference on Uncertainty in Artificial Intelligence}} (Montreal, Quebec, Canada) \emph{(\bibinfo{series}{UAI '09})}. \bibinfo{publisher}{AUAI Press}, \bibinfo{address}{Arlington, Virginia, USA}, \bibinfo{pages}{452–461}.
\newblock
\showISBNx{9780974903958}


\bibitem[Singh et~al\mbox{.}(2024)]%
        {SID2}
\bibfield{author}{\bibinfo{person}{Anima Singh}, \bibinfo{person}{Trung Vu}, \bibinfo{person}{Nikhil Mehta}, \bibinfo{person}{Raghunandan Keshavan}, \bibinfo{person}{Maheswaran Sathiamoorthy}, \bibinfo{person}{Yilin Zheng}, \bibinfo{person}{Lichan Hong}, \bibinfo{person}{Lukasz Heldt}, \bibinfo{person}{Li Wei}, \bibinfo{person}{Devansh Tandon}, \bibinfo{person}{Ed Chi}, {and} \bibinfo{person}{Xinyang Yi}.} \bibinfo{year}{2024}\natexlab{}.
\newblock \showarticletitle{Better Generalization with Semantic IDs: A Case Study in Ranking for Recommendations}. In \bibinfo{booktitle}{\emph{Proceedings of the 18th ACM Conference on Recommender Systems}} (Bari, Italy) \emph{(\bibinfo{series}{RecSys '24})}. \bibinfo{publisher}{Association for Computing Machinery}, \bibinfo{address}{New York, NY, USA}, \bibinfo{pages}{1039–1044}.
\newblock
\showISBNx{9798400705052}
\urldef\tempurl%
\url{https://doi.org/10.1145/3640457.3688190}
\showDOI{\tempurl}


\bibitem[Su et~al\mbox{.}(2021)]%
        {GMCF}
\bibfield{author}{\bibinfo{person}{Yixin Su}, \bibinfo{person}{Rui Zhang}, \bibinfo{person}{Sarah M.~Erfani}, {and} \bibinfo{person}{Junhao Gan}.} \bibinfo{year}{2021}\natexlab{}.
\newblock \showarticletitle{Neural Graph Matching based Collaborative Filtering}. In \bibinfo{booktitle}{\emph{Proceedings of the 44th International ACM SIGIR Conference on Research and Development in Information Retrieval}} (Virtual Event, Canada) \emph{(\bibinfo{series}{SIGIR '21})}. \bibinfo{publisher}{Association for Computing Machinery}, \bibinfo{address}{New York, NY, USA}, \bibinfo{pages}{849–858}.
\newblock
\showISBNx{9781450380379}
\urldef\tempurl%
\url{https://doi.org/10.1145/3404835.3462833}
\showDOI{\tempurl}


\bibitem[Sun et~al\mbox{.}(2019)]%
        {Bert4Rec}
\bibfield{author}{\bibinfo{person}{Fei Sun}, \bibinfo{person}{Jun Liu}, \bibinfo{person}{Jian Wu}, \bibinfo{person}{Changhua Pei}, \bibinfo{person}{Xiao Lin}, \bibinfo{person}{Wenwu Ou}, {and} \bibinfo{person}{Peng Jiang}.} \bibinfo{year}{2019}\natexlab{}.
\newblock \showarticletitle{BERT4Rec: Sequential Recommendation with Bidirectional Encoder Representations from Transformer}. In \bibinfo{booktitle}{\emph{Proceedings of the 28th ACM International Conference on Information and Knowledge Management}} (Beijing, China) \emph{(\bibinfo{series}{CIKM '19})}. \bibinfo{publisher}{Association for Computing Machinery}, \bibinfo{address}{New York, NY, USA}, \bibinfo{pages}{1441–1450}.
\newblock
\showISBNx{9781450369763}
\urldef\tempurl%
\url{https://doi.org/10.1145/3357384.3357895}
\showDOI{\tempurl}


\bibitem[Tang and Wang(2018)]%
        {CASER}
\bibfield{author}{\bibinfo{person}{Jiaxi Tang} {and} \bibinfo{person}{Ke Wang}.} \bibinfo{year}{2018}\natexlab{}.
\newblock \showarticletitle{Personalized Top-N Sequential Recommendation via Convolutional Sequence Embedding}. In \bibinfo{booktitle}{\emph{Proceedings of the Eleventh ACM International Conference on Web Search and Data Mining}} (Marina Del Rey, CA, USA) \emph{(\bibinfo{series}{WSDM '18})}. \bibinfo{publisher}{Association for Computing Machinery}, \bibinfo{address}{New York, NY, USA}, \bibinfo{pages}{565–573}.
\newblock
\showISBNx{9781450355810}
\urldef\tempurl%
\url{https://doi.org/10.1145/3159652.3159656}
\showDOI{\tempurl}


\bibitem[van~den Oord et~al\mbox{.}(2018)]%
        {VQ-VAE}
\bibfield{author}{\bibinfo{person}{Aaron van~den Oord}, \bibinfo{person}{Oriol Vinyals}, {and} \bibinfo{person}{Koray Kavukcuoglu}.} \bibinfo{year}{2018}\natexlab{}.
\newblock \bibinfo{title}{Neural Discrete Representation Learning}.
\newblock
\newblock
\showeprint[arxiv]{1711.00937}~[cs.LG]
\urldef\tempurl%
\url{https://arxiv.org/abs/1711.00937}
\showURL{%
\tempurl}


\bibitem[Vaswani et~al\mbox{.}(2017)]%
        {attention}
\bibfield{author}{\bibinfo{person}{Ashish Vaswani}, \bibinfo{person}{Noam Shazeer}, \bibinfo{person}{Niki Parmar}, \bibinfo{person}{Jakob Uszkoreit}, \bibinfo{person}{Llion Jones}, \bibinfo{person}{Aidan~N. Gomez}, \bibinfo{person}{\L{}ukasz Kaiser}, {and} \bibinfo{person}{Illia Polosukhin}.} \bibinfo{year}{2017}\natexlab{}.
\newblock \showarticletitle{Attention is all you need}. In \bibinfo{booktitle}{\emph{Proceedings of the 31st International Conference on Neural Information Processing Systems}} (Long Beach, California, USA) \emph{(\bibinfo{series}{NIPS'17})}. \bibinfo{publisher}{Curran Associates Inc.}, \bibinfo{address}{Red Hook, NY, USA}, \bibinfo{pages}{6000–6010}.
\newblock
\showISBNx{9781510860964}


\bibitem[Wang et~al\mbox{.}(2019)]%
        {NGCF}
\bibfield{author}{\bibinfo{person}{Xiang Wang}, \bibinfo{person}{Xiangnan He}, \bibinfo{person}{Meng Wang}, \bibinfo{person}{Fuli Feng}, {and} \bibinfo{person}{Tat-Seng Chua}.} \bibinfo{year}{2019}\natexlab{}.
\newblock \showarticletitle{Neural Graph Collaborative Filtering}. In \bibinfo{booktitle}{\emph{Proceedings of the 42nd International ACM SIGIR Conference on Research and Development in Information Retrieval}} \emph{(\bibinfo{series}{SIGIR ’19})}. \bibinfo{publisher}{ACM}, \bibinfo{pages}{165–174}.
\newblock
\urldef\tempurl%
\url{https://doi.org/10.1145/3331184.3331267}
\showDOI{\tempurl}


\bibitem[Wang et~al\mbox{.}(2013)]%
        {NDCG}
\bibfield{author}{\bibinfo{person}{Yining Wang}, \bibinfo{person}{Liwei Wang}, \bibinfo{person}{Yuanzhi Li}, \bibinfo{person}{Di He}, \bibinfo{person}{Tie-Yan Liu}, {and} \bibinfo{person}{Wei Chen}.} \bibinfo{year}{2013}\natexlab{}.
\newblock \bibinfo{title}{A Theoretical Analysis of NDCG Type Ranking Measures}.
\newblock
\newblock
\showeprint[arxiv]{1304.6480}~[cs.LG]


\bibitem[Wei et~al\mbox{.}(2023)]%
        {MGL}
\bibfield{author}{\bibinfo{person}{Chunyu Wei}, \bibinfo{person}{Jian Liang}, \bibinfo{person}{Di Liu}, \bibinfo{person}{Zehui Dai}, \bibinfo{person}{Mang Li}, {and} \bibinfo{person}{Fei Wang}.} \bibinfo{year}{2023}\natexlab{}.
\newblock \showarticletitle{Meta Graph Learning for Long-tail Recommendation}. In \bibinfo{booktitle}{\emph{Proceedings of the 29th ACM SIGKDD Conference on Knowledge Discovery and Data Mining}} (Long Beach, CA, USA) \emph{(\bibinfo{series}{KDD '23})}. \bibinfo{publisher}{Association for Computing Machinery}, \bibinfo{address}{New York, NY, USA}, \bibinfo{pages}{2512–2522}.
\newblock
\showISBNx{9798400701030}
\urldef\tempurl%
\url{https://doi.org/10.1145/3580305.3599428}
\showDOI{\tempurl}


\bibitem[Wu et~al\mbox{.}(2019)]%
        {SRGNN}
\bibfield{author}{\bibinfo{person}{Shu Wu}, \bibinfo{person}{Yuyuan Tang}, \bibinfo{person}{Yanqiao Zhu}, \bibinfo{person}{Liang Wang}, \bibinfo{person}{Xing Xie}, {and} \bibinfo{person}{Tieniu Tan}.} \bibinfo{year}{2019}\natexlab{}.
\newblock \showarticletitle{Session-Based Recommendation with Graph Neural Networks}.
\newblock \bibinfo{journal}{\emph{Proceedings of the AAAI Conference on Artificial Intelligence}} \bibinfo{volume}{33}, \bibinfo{number}{01} (\bibinfo{date}{Jul.} \bibinfo{year}{2019}), \bibinfo{pages}{346--353}.
\newblock
\urldef\tempurl%
\url{https://doi.org/10.1609/aaai.v33i01.3301346}
\showDOI{\tempurl}


\bibitem[Xia et~al\mbox{.}(2023)]%
        {SimRec}
\bibfield{author}{\bibinfo{person}{Lianghao Xia}, \bibinfo{person}{Chao Huang}, \bibinfo{person}{Jiao Shi}, {and} \bibinfo{person}{Yong Xu}.} \bibinfo{year}{2023}\natexlab{}.
\newblock \showarticletitle{Graph-less Collaborative Filtering}. In \bibinfo{booktitle}{\emph{Proceedings of the ACM Web Conference 2023}} (Austin, TX, USA) \emph{(\bibinfo{series}{WWW '23})}. \bibinfo{publisher}{Association for Computing Machinery}, \bibinfo{address}{New York, NY, USA}, \bibinfo{pages}{17–27}.
\newblock
\showISBNx{9781450394161}
\urldef\tempurl%
\url{https://doi.org/10.1145/3543507.3583196}
\showDOI{\tempurl}


\bibitem[Xia et~al\mbox{.}(2022)]%
        {HCCF}
\bibfield{author}{\bibinfo{person}{Lianghao Xia}, \bibinfo{person}{Chao Huang}, \bibinfo{person}{Yong Xu}, \bibinfo{person}{Jiashu Zhao}, \bibinfo{person}{Dawei Yin}, {and} \bibinfo{person}{Jimmy Huang}.} \bibinfo{year}{2022}\natexlab{}.
\newblock \showarticletitle{Hypergraph Contrastive Collaborative Filtering}. In \bibinfo{booktitle}{\emph{Proceedings of the 45th International ACM SIGIR Conference on Research and Development in Information Retrieval}} (Madrid, Spain) \emph{(\bibinfo{series}{SIGIR '22})}. \bibinfo{publisher}{Association for Computing Machinery}, \bibinfo{address}{New York, NY, USA}, \bibinfo{pages}{70–79}.
\newblock
\showISBNx{9781450387323}
\urldef\tempurl%
\url{https://doi.org/10.1145/3477495.3532058}
\showDOI{\tempurl}


\bibitem[Xie et~al\mbox{.}(2022)]%
        {CL4SRec}
\bibfield{author}{\bibinfo{person}{Xu Xie}, \bibinfo{person}{Fei Sun}, \bibinfo{person}{Zhaoyang Liu}, \bibinfo{person}{Shiwen Wu}, \bibinfo{person}{Jinyang Gao}, \bibinfo{person}{Jiandong Zhang}, \bibinfo{person}{Bolin Ding}, {and} \bibinfo{person}{Bin Cui}.} \bibinfo{year}{2022}\natexlab{}.
\newblock \showarticletitle{Contrastive Learning for Sequential Recommendation}. In \bibinfo{booktitle}{\emph{2022 IEEE 38th International Conference on Data Engineering (ICDE)}}. \bibinfo{pages}{1259--1273}.
\newblock
\urldef\tempurl%
\url{https://doi.org/10.1109/ICDE53745.2022.00099}
\showDOI{\tempurl}


\bibitem[Ye et~al\mbox{.}(2023)]%
        {MAERec}
\bibfield{author}{\bibinfo{person}{Yaowen Ye}, \bibinfo{person}{Lianghao Xia}, {and} \bibinfo{person}{Chao Huang}.} \bibinfo{year}{2023}\natexlab{}.
\newblock \showarticletitle{Graph Masked Autoencoder for Sequential Recommendation}. In \bibinfo{booktitle}{\emph{Proceedings of the 46th International ACM SIGIR Conference on Research and Development in Information Retrieval}} (Taipei, Taiwan) \emph{(\bibinfo{series}{SIGIR '23})}. \bibinfo{publisher}{Association for Computing Machinery}, \bibinfo{address}{New York, NY, USA}, \bibinfo{pages}{321–330}.
\newblock
\showISBNx{9781450394086}
\urldef\tempurl%
\url{https://doi.org/10.1145/3539618.3591692}
\showDOI{\tempurl}


\bibitem[Zhang et~al\mbox{.}(2019)]%
        {StarGCN}
\bibfield{author}{\bibinfo{person}{Jiani Zhang}, \bibinfo{person}{Xingjian Shi}, \bibinfo{person}{Shenglin Zhao}, {and} \bibinfo{person}{Irwin King}.} \bibinfo{year}{2019}\natexlab{}.
\newblock \showarticletitle{STAR-GCN: stacked and reconstructed graph convolutional networks for recommender systems}. In \bibinfo{booktitle}{\emph{Proceedings of the 28th International Joint Conference on Artificial Intelligence}} (Macao, China) \emph{(\bibinfo{series}{IJCAI'19})}. \bibinfo{publisher}{AAAI Press}, \bibinfo{pages}{4264–4270}.
\newblock
\showISBNx{9780999241141}


\bibitem[Zheng et~al\mbox{.}(2024a)]%
        {LCREC}
\bibfield{author}{\bibinfo{person}{Bowen Zheng}, \bibinfo{person}{Yupeng Hou}, \bibinfo{person}{Hongyu Lu}, \bibinfo{person}{Yu Chen}, \bibinfo{person}{Wayne~Xin Zhao}, \bibinfo{person}{Ming Chen}, {and} \bibinfo{person}{Ji-Rong Wen}.} \bibinfo{year}{2024}\natexlab{a}.
\newblock \showarticletitle{Adapting Large Language Models by Integrating Collaborative Semantics for Recommendation}. In \bibinfo{booktitle}{\emph{2024 IEEE 40th International Conference on Data Engineering (ICDE)}}. \bibinfo{pages}{1435--1448}.
\newblock
\urldef\tempurl%
\url{https://doi.org/10.1109/ICDE60146.2024.00118}
\showDOI{\tempurl}


\bibitem[Zheng et~al\mbox{.}(2024b)]%
        {LC-REC}
\bibfield{author}{\bibinfo{person}{Bowen Zheng}, \bibinfo{person}{Yupeng Hou}, \bibinfo{person}{Hongyu Lu}, \bibinfo{person}{Yu Chen}, \bibinfo{person}{Wayne~Xin Zhao}, \bibinfo{person}{Ming Chen}, {and} \bibinfo{person}{Ji-Rong Wen}.} \bibinfo{year}{2024}\natexlab{b}.
\newblock \showarticletitle{Adapting Large Language Models by Integrating Collaborative Semantics for Recommendation}. In \bibinfo{booktitle}{\emph{2024 IEEE 40th International Conference on Data Engineering (ICDE)}}. \bibinfo{pages}{1435--1448}.
\newblock
\urldef\tempurl%
\url{https://doi.org/10.1109/ICDE60146.2024.00118}
\showDOI{\tempurl}


\bibitem[Zheng et~al\mbox{.}(2019)]%
        {Sparsity}
\bibfield{author}{\bibinfo{person}{Lei Zheng}, \bibinfo{person}{Chaozhuo Li}, \bibinfo{person}{Chun-Ta Lu}, \bibinfo{person}{Jiawei Zhang}, {and} \bibinfo{person}{Philip~S. Yu}.} \bibinfo{year}{2019}\natexlab{}.
\newblock \showarticletitle{Deep Distribution Network: Addressing the Data Sparsity Issue for Top-N Recommendation}. In \bibinfo{booktitle}{\emph{Proceedings of the 42nd International ACM SIGIR Conference on Research and Development in Information Retrieval}} (Paris, France) \emph{(\bibinfo{series}{SIGIR'19})}. \bibinfo{publisher}{Association for Computing Machinery}, \bibinfo{address}{New York, NY, USA}, \bibinfo{pages}{1081–1084}.
\newblock
\showISBNx{9781450361729}
\urldef\tempurl%
\url{https://doi.org/10.1145/3331184.3331330}
\showDOI{\tempurl}


\bibitem[Zhu et~al\mbox{.}(2020)]%
        {BGNN}
\bibfield{author}{\bibinfo{person}{Hongmin Zhu}, \bibinfo{person}{Fuli Feng}, \bibinfo{person}{Xiangnan He}, \bibinfo{person}{Xiang Wang}, \bibinfo{person}{Yan Li}, \bibinfo{person}{Kai Zheng}, {and} \bibinfo{person}{Yongdong Zhang}.} \bibinfo{year}{2020}\natexlab{}.
\newblock \showarticletitle{Bilinear graph neural network with neighbor interactions}.
\newblock \bibinfo{journal}{\emph{arXiv preprint arXiv:2002.03575}} (\bibinfo{year}{2020}).
\newblock


\end{thebibliography}

\end{document}